\documentclass[%
reprint,
showpacs,preprintnumbers,
amsmath,amssymb,
aps,
prc,
]{revtex4-1}
\usepackage{graphicx}% Include figure files
\usepackage{bm}% bold math
\usepackage{epstopdf}
\newcommand{\cc}{\mathrm{c\bar{c}}}

\newcommand{\bb}{\mathrm{b\bar{b}}}

\newcommand{\Npart}{{\langle N_\text{part} \rangle}}

\usepackage{amsmath}
\usepackage{bm}
\usepackage{mathtools}
\usepackage{bm}
\usepackage{multirow}

\usepackage[ngerman,english]{babel}
\usepackage{mathtools}

\usepackage{bm}% bold math
\usepackage{epstopdf}
\usepackage{amsbsy}
\usepackage{amsmath}
\usepackage{amssymb}
\usepackage[separate-uncertainty = true,multi-part-units=single]{siunitx}
\usepackage{commath}
\usepackage{graphicx}

\newcommand{\textoverline}[1]{$\overline{\mbox{#1}}$}
\begin{document}

%\preprint{APS/123-QED}
\title{Aspects of relativistic heavy-ion collisions}% Force line breaks with \\
%\thanks{A footnote to the article title}
\author{Georg Wolschin}
\email{g.wolschin@thphys.uni-heidelberg.de}
\affiliation{Institut f{\"ur} Theoretische Physik der Universit{\"a}t Heidelberg, Philosophenweg 12-16, D-69120 Heidelberg, Germany, EU}% Authors' institution and/or address
%\collaboration{MUSO Collaboration}
%\noaffiliation
%\date{\today}% It is always \today, today, but any date may be explicitly specified
\begin{abstract}
The rapid thermalization of quarks and gluons in the initial stages of relativistic heavy-ion collisions is treated using analytic solutions of a nonlinear diffusion equation with
schematic initial conditions, and for gluons with boundary conditions at the singularity. On a similarly short time scale of $t \le1$ fm/$c$, the stopping of baryons is accounted for through a QCD-inspired approach based on the parton distribution functions of valence quarks, and gluons. Charged-hadron production is considered phenomenologically using a linear relativistic diffusion model with two fragmentation sources, and a central gluonic source that rises with $\ln^3(s_{NN})$. The limiting-fragmentation conjecture that agrees with data at energies reached at the Relativistic Heavy Ion Collider (RHIC) is found to be consistent with Large Hadron Collider (LHC) data for Pb-Pb at $\sqrt{s_{NN}}= 2.76$ and $5.02$ TeV. Quarkonia are used as hard probes for the properties of the quark-gluon plasma (QGP) through a comparison of theoretical predictions with recent CMS, ALICE and LHCb data for Pb-Pb and p-Pb collisions.
\end{abstract}
\pacs{25.75.-q,25.75.Dw,25.75.C}% PACS, the Physics and Astronomy Classification Scheme.
%\keywords{Suggested keywords}% Use showkeys class option if keyword display desired
\maketitle
\section{Introduction}
%\label{sec:intro}%\label{}
This article covers aspects of relativistic heavy-ion collisions in the energy regions reached at the Relativistic Heavy Ion Collider (RHIC) and the Large Hadron Collider (LHC). Starting with the thermalization of partons in the initial stages, stopping of the incoming baryons is discussed, followed by charged-hadon production and the modification of quarkonia in the hot quark-gluon plasma (QGP), with an emphasis on bottomonia that provide clear signals of QGP formation. The overall approach is phenomenological and in close comparison with data. In some cases such as stopping and quarkonia, nonequilibrium-statistical considerations are merged with quantum chromodynamics (QCD). The work contains both new developments and re-views of our previously published work in a new context.

The fast local thermalization of partons in the initial stages of relativistic heavy-ion collisions is a sufficient condition to apply hydrodynamic descriptions \cite{hesne13} of the subsequent collective expansion and cooling of the hot fireball that is created in the collision. Typical local equilibration times for gluons are about 0.1 fm/$c$ \cite{fmr18}, with initial central temperatures of the order of 500--600 MeV reached in a Pb-Pb collision at $\sqrt{s_\text{NN}}=5.02$ TeV at the Large Hadron Collider \cite{hnw17}. Thermalization times for quarks are typically by an order of magnitude larger \cite{gw18} due to the smaller color factor and the different statistical properties (Pauli's principle). Whereas the thermal Bose--Einstein/Fermi--Dirac distributions and also the initial distribution for quarks are known, plausible assumptions are being made for the primordial gluon distribution before the onset of the collision.

Studies such as Ref.\,\cite{he18} have found that local thermalization may not be a necessary condition for the applicability of hydrodynamics to relativistic heavy-ion collisions. This would imply that hydrodynamics could be a valid approach away from local equilibrium \cite{ro17} -- but still, it remains important to investigate how and on which timescale local thermalization is achieved. 
Obviously,~an~investigation of the local thermalization of quarks and gluons makes 
sense only in a weakly coupled description based on an effective kinetic theory 
that relies on the Boltzmann equation. It is recognized that a strongly coupled 
paradigm built on the anti-de Sitter-conformal field theory (AdS/CFT) %please define
 correspondence \cite{son07} discovered in the investigation 
of D-branes in string theory \cite{mal98} may be relevant at RHIC % please define
and LHC energy scales 
 such that partons would not be the relevant degrees 
of freedom any more. In~this work, however, I rather assume quarks and gluons 
to be well-defined and long-lived excitations in QCD at temperatures close 
to the critical value.

What is not known precisely is the development with time towards local equilibrium. An easy way to approximate the time evolution is given by the linear relaxation-time approximation (RTA), which provides a simple analytic solution of the problem by enforcing a linear transition from the initial to the local equilibrium state, but does not properly account for the known nonlinearity of the system. A more ambitious approach for bosons is to numerically solve gluon transport equations that include the effect of Bose statistics, as has been done in Ref.\,\cite{jpb12} and subsequent works. Initially, only elastic scattering was considered with the possibility of gluon condensate formation due to particle-number conservation in over-populated systems, but it was recognized that inelastic, number-changing radiative processes cannot be neglected \cite{blmt17}, and hinder the formation of a Bose condensate.

Whereas also other numerical calculations relying on a quantum Boltzmann collision term to account for the initial local equilibration are available such as Ref.\,\cite{xu15}, it is of interest to have an exactly solvable analytic model to better understand the physics of the fast equilibration. A corresponding nonlinear boson diffusion equation (NBDE) has been presented in Ref.\,\cite{gw18} and solved for a simplified case that did, however, not yet consider the singularity at the chemical potential $\mu<0$. The nonlinear partial differential equation preserves the essential features of Bose--Einstein statistics that are contained in the collision term. In particular, the thermal equilibrium distribution emerges as a stationary solution and hence, the equation appears suitable to model the thermalization of gluons in relativistic collisions. It is used in this work to generate new exact solutions for the time-dependent gluon equilibration problem that include boundary conditions at the singularity. Regarding fermionic thermalization, the corresponding nonlinear fermion diffusion equation is easier to solve \cite{gw18,bgw19} because no singularity appears, and the analytic solutions will be reviewed.

On an equally short time scale as the local equilibration, the incoming baryons with energies available at RHIC or LHC are \it{stopped:} \rm The system is slowing down, essentially through collisions of the incoming valence quarks with soft gluons in the respective other nucleus. Various models to account for this process and its energy dependence have been developed, in particular in Refs.\,\cite{mtw09,mtwc09} and related works, which are relying on the appropriate parton distribution functions (PDFs) and hence, on QCD, yielding good agreement with the available net-proton (proton minus antiproton) stopping data. Different from the nonequilibrium-statistical approach to initial thermalization, such models do not consider a time dependence. However, by using the rapidity distribution calculated from the PDFs and the initial valence-quark distribution, one can, in addition, account for the time development from the initial to the final distribution with an appropriate fluctuation--dissipation relation.

The relevant bulk properties of relativistic heavy-ion collisions mostly arise from charged-hadron production. In many of the available macroscopic and microscopic models, hadronization occurs from the fireball at the phase boundary between the QGP and the hadronic phase. However, it has been proposed \cite{gw13,gw16} that the production of charged hadrons from the fragmentation sources at larger rapidity values is also relevant in the overall distributions and should be treated separately from the fireball source. At midrapidity, these contributions are relatively small, but become relevant more forward or backward. Corresponding new results of the phenomenological three-source relativistic diffusion model (RDM) are compared with recent LHC pseudorapidity data on charged-hadron production in 5.02 TeV Pb-Pb, which are also shown to be consistent with the limiting-fragmentation hypothesis that had been found to agree with the hadron production data at RHIC energies \cite{bea02,bb03,ada06}.

Regarding more direct evidence for transient QGP formation in relativistic heavy-ion collisions, jets may provide the most direct manifestation for quarks and gluons in the system. In particular, the suppression of away-side jets in the hot QGP had already been predicted by Bjorken \cite{bj82}, and confirmed experimentally by the STAR collaboration \cite{star04}. Meanwhile, jet suppression at LHC energies has been investigated in detail (e.\,g.~Ref.\,\cite{cms11}), and it has been shown how strong final-state interactions cause high-$p_\text{T}$ jets to lose energy to the plasma.

In this work, however, I consider quarkonia as another indicator for the properties of the quark-gluon plasma such as its initial central temperature $T_\text{i}$. Quarkonia are bound states of heavy quark-antiquark pairs that can be formed in initial hard partonic collisions.
In the original prediction for a suppression of the $J/\psi$ yields in the presence of a QGP, only the medium-effect on the real part of the quark-antiquark potential was considered \cite{ms86}. It has later been realized that due to the presence of the hot medium, the potential has an imaginary part \cite{laine07}. Optical potentials had also been used in the theory of nuclear reactions to account for channels that are not treated explicitly.  In case of quarkonia, the imaginary part causes their \textit{dissociation}, in addition to melting of the quarkonia states because the real potential is screened.
It is possible to treat quarkonia dissociation by thermal gluons separately from the imaginary part \cite{bgw12}.

In case of charmonium at LHC energies, statistical recombination of charm and anticharm quarks turns out to be important, but it is not possible to separate the process from the dissociation in the QGP. One may therefore concentrate on the heavier bottomonium system, where recombination is much less pronounced. Through detailed investigations of the transverse-momentum- and centrality-dependent suppression of the spin-triplet $\Upsilon(1S,2S)$ states in Pb-Pb collisions at LHC energies and comparisons with CMS data, we can deduce QGP properties such as the initial central temperature $T_\text{i}$ and study its dependence on the center-of-mass energy.

In asymmetric collisions such as p-Pb, the situation is quite different from symmetric systems, because most of the system remains cold due to the much smaller overlap. Cold nuclear matter (CNM) effects therefore provide a certain understanding of the measured quarkonia modifications, but a complete agreement with the available data remains impossible -- unless one also considers the hot QGP zone that is still produced, even though it is initially considerably less extended as compared to symmetric systems. Indeed the bottomonia dissociation in the hot QGP provides a clue for the interpretation of the data in asymmetric collisions as well.

The paper follows the above-mentioned series of topics: In Section 2, the nonlinear diffusion equation for gluons and quarks is solved explicitly to account for the fast local thermalization at the beginning of relativistic heavy-ion collisions. Stopping is considered in Section 3, hadron production and limiting fragmentation in Section 4, bottomonia modification in the medium in Section 5. The conclusions are drawn in Section 6.

\section{Fast Thermalization of Gluons and Quarks: An Analytic Nonlinear Model}
For a given initial nonequilibrium gluon distribution at $t=0$, solutions of the nonlinear boson diffusion equation describe the time-dependent equilibration towards the thermal distribution with the local temperature $T$. In Ref.\,\cite{gw18} such solutions were calculated with the free Green's function. Whereas this accounts for local thermalization in the ultraviolet (UV) with the corresponding equilibration time $\tau_\text{eq}$, in the infrared (IR) the populations decrease due to diffusion into the negative-energy region.
To avoid such an unphysical behaviour, one has to consider the boundary condition at the singularity $|\bf{p} \rm|=p=$ $\epsilon=\mu$ with the chemical potential $\mu<0$, and the corresponding bounded Green's function in the solution of the NBDE. With this Green's function, gluon populations indeed attain the Bose--Einstein limit also in the infrared for nonequilibrium initial conditions that include the singularity.

The nonlinear model and the solution of the combined initial- and boundary-value problem are first briefly reviewed. Subsequently, the thermalization problem is solved for a schematic initial gluon distribution that characterizes the relativistic collision at $t=0$. Adding the boundary condition at the singularity, $n(\epsilon=\mu<0,t)\rightarrow\infty$, the time-dependent partition function that includes initial and boundary conditions is obtained using analytic expressions for both, the bound Green's function, and the function that contains an integral over the initial conditions. The resulting occupation-number distribution function $n(\epsilon,t)$ is calculated, and it is shown to approach the equilibrium distribution both in the UV and in the IR.

%The rapid thermalization of gluons and quarks in the initial stages of relativistic heavy-ion collisions will be treated using analytic solutions of a nonlinear diffusion equation with
%schematic initial conditions, and for gluons with boundary conditions at the singularity $\epsilon=\mu<0$.
%The solutions describe the time-dependent approach of gluons to the Bose--Einstein equilibrium distribution with a local equilibration time of $\tau_\text{eq}\simeq 0.1$ fm/$c$ and %central temperatures of the order of $500-600$ MeV in the initial stages of Pb-Pb collisions at energies reached at the Large Hadron Collider (LHC), and of quarks to the Fermi-%Dirac distribution \cite{gw18}.
Before we proceed to the nonlinear model, it is useful to consider a linear time-dependent transition from the initial distribution
%Before proceeding to the nonlinear model, it is useful to consider a linear time-dependent relaxation from the initial distribution
\begin{equation}
	%n_\text{i}(\epsilon)=N_\text{i}\,\theta(1-\epsilon/Q_\text{s})\,\theta(\epsilon)\,.
	n_\text{i}(\epsilon)=\theta(1-\epsilon/Q_\text{s})\,\theta(\epsilon)
	\label{nini}
\end{equation}
%--which will be discussed later in more detail--
to the thermal distribution
% to the thermal distribution $n_\text{eq}(\epsilon)$
\begin{equation}
	n_\text{eq}(\epsilon)=\frac{1}{e^{(\epsilon-\mu)/T}-1}
	\label{Bose-Einstein}
\end{equation}
with the chemical potential $\mu<0$ in a finite boson system in the relaxation-time approximation (RTA), $\partial\,n_\text{rel}/\partial t=(n_\text{eq} -n_\text{rel})/\tau_\text{eq}$, with solution
\begin{equation}
	%n\,(\epsilon,t)=n_\text{i}(\epsilon)\exp(-t/\tau_{eq})+n_\text{eq}(\epsilon)(1-\exp(-t/\tau_{eq}))\,.
	n_\text{rel}(\epsilon,t)=n_\text{i}(\epsilon)\,e^{-t/\tau_\text{eq}}+n_\text{eq}(\epsilon)(1-e^{-t/\tau_\text{eq}})\,.
	\label{rela}
\end{equation}
\begin{figure}[t]
	%\begin{widetext}
	\centering
	\includegraphics[scale=0.42]{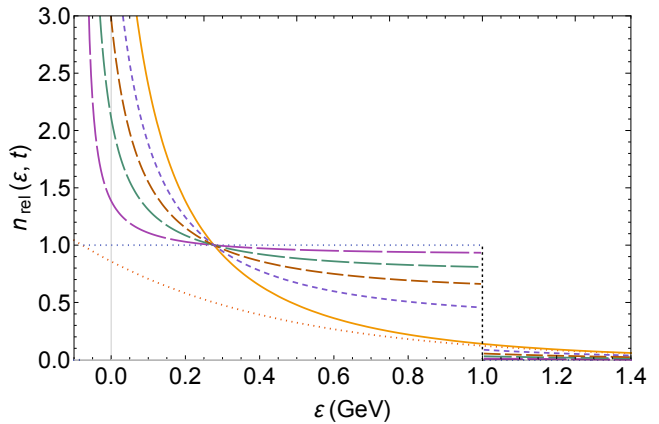}% Here is how to import EPS art
	\caption{\label{fig1}
	Local thermalization of gluons in the linear relaxation-time approximation (RTA) for $\mu<0$. Starting from schematic initial conditions Eq.\,(\ref{inix}) in the cold system at $t=0$ (box distribution with cut at $\epsilon=Q_\text{s}=1$ GeV), a Bose--Einstein equilibrium distribution with temperature $T\simeq513$ MeV (solid curve) is approached. Time-dependent single-particle occupation-number distribution functions are shown at $t = 0.02$, $0.08$, $0.15$, $0.3$, and $0.6$ fm/$c$ (decreasing dash lenghts). The lower dotted curve is Boltzmann's distribution.
	%Thermalization occurs much slower than in the nonlinear case, see Figure\,\ref{fig1}.
	}
	%The equilibrium distribution at temperature $T_\text{f}$ is reached within $t\simeq 14$ ms.
	%\end{widetext}
\end{figure}
Time-dependent RTA results for gluons are shown in Fig.\,\ref{fig1} for $t = 0.02$, $0.08$, $0.15$, $0.3$, and $0.6$ fm/$c$. The thermal distribution with initial central temperature $T=513$ MeV as inferred in Ref.\,\cite{hnw17} for central Pb-Pb collisions at $\sqrt{s_{NN}}=5.02$ TeV is approached linearly, the discontinuities at $\epsilon=Q_\text{s}$ persist.
%Thermalization for fixed $\tau_\text{eq}$ occurs relatively slowly.

For a more realistic account of thermalization, one needs to consider the inherent nonlinearity of the system.
 I had derived a  nonlinear partial differential equation
 for the single-particle occupation probability distributions $n\equiv n_\text{th}(\epsilon,t)$ 
 from the bosonic/fermionic Boltzmann collision term in Ref.\,\cite{gw18}. 
 The transport coefficients in this nonlinear boson diffusion equation (NBDE) depend on energy, 
 time, and the second moment of the interaction. They incorporate the complicated many-body physics. The drift term $v(\epsilon,t)$ is responsible for 
 dissipative effects, the diffusion term $D(\epsilon,t)$ for diffusion of particles in the energy 
 space. For the simplified case of energy-independent transport coefficients, the nonlinear diffusion 
 equation for the occupation-number distribution $n^\pm (\epsilon,t)$ becomes

\begin{equation}
	\frac{\partial n^\pm}{\partial t}=-v\,\frac{\partial}{\partial\epsilon}\Bigl[n\,(1\pm n)\Bigr]+D\,\frac{\partial^2n}{\partial\epsilon^2}
	\label{bose}
\end{equation}
where the $+$ sign represents bosons, and the $-$ sign fermions.
A stationary solution is given by the thermal distribution Eq.\,(\ref{Bose-Einstein}) for bosons and, correspondingly, the Fermi--Dirac distribution for fermions (quarks).
In spite of its simple structure, the nonlinear diffusion equation with constant transport coefficients preserves the essential features of Bose--Einstein/Fermi--Dirac statistics which are contained in the quantum Boltzmann equation.

The transport equation can be solved exactly for a given initial condition $n^\pm_\text{i}(\epsilon)$ using the nonlinear transformation outlined in Ref.\,\cite{gw18}.
For gluons, the nonlinear boson diffusion equation (NBDE) is more difficult to solve analytically due to the singularity at the chemical potential $\epsilon=\mu<0$, and the need to consider the boundary conditions at the singularity. For fermions, there is no corresponding singularity, such that the exact solution of the nonlinear problem can be obtained with the free Green's function as performed in Refs.\,\cite{gw18,bgw19}. For gluons, the bounded solution $n^+(\epsilon,t)$
%\equiv n(\epsilon,t)$
\cite{gw20a} is briefly reviewed. It can be written as
\begin{align}
	n^+(\epsilon,t) = -\frac{D}{v} \frac{\partial}{\partial\epsilon}\ln{\mathcal{Z}^+(\epsilon,t)} -\frac{1}{2}= -\frac{D}{v}\frac{1}{\mathcal{Z}^+} \frac{\partial\mathcal{Z}^+}{\partial\epsilon} -\frac{1}{2}
	\label{eq:Nformula}
\end{align}
with the time-dependent partition function $\mathcal{Z}^+(\epsilon,t)\equiv \mathcal{Z}(\epsilon,t)$ obeying a linear diffusion equation
\begin{align}
	\frac{\partial}{\partial t}{\mathcal{Z}}(\epsilon,t) = D \frac{\partial^2}{\partial\epsilon^2}{\mathcal{Z}}(\epsilon,t)\,.
	\label{eq:diffusionequation}
\end{align}
In the absence of boundary conditions, the free partition function becomes
\begin{align}
	\mathcal{Z}_\text{free}(\epsilon,t)= a(t)\int_{-\infty}^{+\infty} G_\text{free}(\epsilon,x,t)\,F(x)\,\text{d}x\,,
	\label{eq:partitionfunctionZ}
\end{align}
The energy-independent prefactor $a(t)$ in the  partition function drops out when taking the logarithmic derivative in the calculation of the occupation-number distribution.
The function $F(x)$  with the integral over the initial conditions covers the full energy region $-\infty<x<\infty$. Due to the occurence of the singularity, however,  one eventually will have to consider boundary conditions at $\epsilon=\mu<0$.

For a solution without boundary conditions as in Refs.\,\cite{gw18}, Green's function \(G_\text{free}(\epsilon , x , t)\) of Eq.\,(\ref{eq:diffusionequation}) is a single Gaussian
\begin{align}
	G_\text{free}(\epsilon,x,t)=\exp\Bigl(- \frac{(\epsilon-x)^2}{4Dt}\Bigr)\,,
	\label{eq:Greensnonfixed}
\end{align}
but it becomes more involved once boundary conditions are considered.
The function \(F(x)\) depends on the initial occupation-number distribution $n_\mathrm{i}$ according to
\begin{align}
	F(x) = \exp\Bigl[ -\frac{1}{2D}\bigl( v x+2v \int_0^x n_{\mathrm{i}}(y)\,\text{d}y \bigr) \Bigr]\,.
	\label{ini}
\end{align}
As discussed in Ref.\, \cite{rgw20}, the definite integral
can be replaced w.l.o.g. by the indefinite integral $A_{\mathrm{i}}(x)$ over the initial distribution with $\partial_x A_{\mathrm{i}} (x) = n_{\mathrm{i}}(y)$, resulting in
\begin{align}
	F(x) = \exp\Bigl[-\frac{1}{2D}\left( v x+2v A_{\mathrm{i}}(x) \right)\Bigr]\,.
	\label{fini}
\end{align}
%This replacement still provides the exact solution.
%It is now possible to compute the partition function and the overall solution for the occupation-number distribution function Eq.\,(\ref{eq:Nformula}) analytically, even in the presence of %a singularity in the initial conditions.
% -- which had been excluded in the initial conditions, and hence, in the solutions given in Refs.\,\cite{gw18,gw18a}.
%This replacement still provides the exact solution.
For any given initial distribution $n_{\mathrm{i}}$, one can now compute the partition function and the overall solution for the occupation-number distribution function Eq.\,(\ref{eq:Nformula}) analytically.
% even in the presence of a singularity in the initial conditions.
%(The singularity had been excluded in the solution given in Ref.\,\cite{gw18a}).
%To obtain physically meaningful solutions not only in the UV, but also in the IR, one has to consider the boundary conditions at the singularity.
The solution technique has been developed in Refs.\,\cite{rgw20,gw20} for the case of a cold bosonic atom gas that undergoes evaporative cooling. Here, and in Ref.\,\cite{gw20a} for different initial conditions, the approach is carried over to equilibrating gluons at relativistic energies.
%To solve the problem with boundary conditions at the singularity,
To solve the problem exactly,
%$n(\epsilon=\mu,t)=\infty$,
the chemical potential is treated as a fixed parameter. With \(\lim_{\epsilon \downarrow \mu} n(\epsilon,t) = \infty\) \,$\forall$ \(t\), one obtains \( \mathcal{Z} (\mu,t) = 0\), and the energy range is restricted to $\epsilon \ge \mu$. This requires a new Green's function
%\( {F} (\epsilon,x,t) \) \cite{EqWorld}
that equals zero at \(\epsilon = \mu\) $\forall \,t$. It can be written as
\begin{align}
	{G} (\epsilon,x,t) = G_\text{free}(\epsilon - \mu,x,t) - G_\text{free}(\epsilon - \mu,-x,t)\,,
	\label{Greens}
\end{align}
and the partition function with this boundary condition becomes
\begin{align}
	{\mathcal{Z}} (\epsilon,t) = \int_0^\infty {G} (\epsilon, x, t)\,F(x+\mu)\, \text{d}x\,.
	\label{eq:newformulaforZ}
\end{align}
%Here, the Green's function Eq.\,\eqref{eq:newGreens} restricts the integral to energies $\epsilon \ge \mu$.
%This gives rise to a new expression for the occupation-number distribution.
%The corresponding solution for the occupation-number distribution with fixed chemical potential is equivalent to imposing a point symmetry around \(\mu\) in the initial distribution $n_\text{i}(\epsilon)$ that appears in \(G(x)\).
The function $F$ remains unaltered with respect to Eq.\,(\ref{fini}), except for a shift of its argument by the chemical potential. With a given initial nonequilibrium distribution $n_\text{i}$, the NBDE can now be solved including boundary conditions at the singularity. The solution is given by Eq.\,(\ref{eq:Nformula}).
\subsection{Thermalization of Gluons}
For massless gluons at the onset of a relativistic hadronic collision, an initial-momentum distribution $n_\text{i}(|\mathbf{p}|)\equiv\,n_\text{i}(p)=n_\text{i}\,(\epsilon)$ has been proposed by Mueller \cite{mue00} based on Ref.\,\cite{mlv94}.
It accounts, in particular, for the situation at the start of a relativistic heavy-ion collision \cite{jpb12}. It amounts to assuming that all gluons up to a limiting momentum $Q_\text{s}$ are freed on a short time scale $\tau_0\sim Q_\text{s}^{-1}$, whereas all gluons beyond $Q_\text{s}$ are not freed.
%for times $t\lesssim 1$ fm/$c\simeq 0.33\times 10^{-23}$ s
Thus the initial gluon-mode occupation in a volume $V$ is taken to be a constant up to $Q_\text{s}$, as in Eq.\,(\ref{nini}) that was already used before in the relaxation-time approximation.
%, and all negative-energy states for bosons are empty
%=Q_0\,A^{1/6}x^{-\lambda/2}$ with the mass number $A$, the longitudinal
%momentum fraction $x$ carried by the gluon and the saturation-scale exponent $\lambda\simeq 0.2-0.3$,
%\begin{equation}
%n_\text{i}(\epsilon)=N_\text{i}\,\theta(1-\epsilon/Q_\text{s})\,\theta(\epsilon)\,.
%n_\text{i}(\epsilon)=\theta(1-\epsilon/Q_\text{s})\,\theta(\epsilon)
%\label{nini}
%\end{equation}
%In the corresponding fermionic case \cite{gw82}, all negative-energy states are filled as in the Dirac sea.
% The momentum scale is set by $Q_0$, and
Typical gluon saturation momenta for a longitudinal momentum fraction carried by the gluon $x\simeq 0.01$ turn out to be of the order $Q_\text{s}\simeq 1$ GeV \cite{mtw09}, which is chosen for the present model investigation.
%\begin{equation}
%\rho(\epsilon)=\int\frac{d^3p}{(2\pi)^3}\delta(\epsilon-|\bm{p}|)\propto \epsilon^2\,.
%\label{density}
%\end{equation}
%such that the integral over the energy gives the total initial gluon number $\int_0^\infty n_\text{i}(\epsilon,t)\rho(\epsilon)=n_\text{i}$.

Results for the gluon thermalization from $n_\text{i}(\epsilon)$
%(with $n_\text{i}=1$)
to $n_\text{eq}(\epsilon)$ according to Eq.\,(\ref{bose}) have been calculated in Ref.\,\cite{gw18} for the free case, without considering boundary conditions at the singularity.
%$\epsilon=\mu<0$.
As a consequence, diffusion into the negative-energy region occured, depleting the occupation in the infrared such that the asymptotic distribution differed from Bose--Einstein.

%with the equilibrium temperature $T$ that is attained upon completion of the local equilibration, and a chemical potential $\mu$ which is adjusted such that the two partial %distributions match at $\epsilon=0$, with $n_\text{i}(0)=1$.

As a remedy, one has to extend the energy scale in Eq.\,(\ref{nini}) to $\mu\le \epsilon<\infty$, and include the boundary conditions at the singularity $\epsilon=\mu<0$. This will cause the time-dependent solutions of the NBDE to properly approach the thermal Bose--Einstein distribution over the full energy scale as $t\rightarrow \infty$. The initial condition is thus modified to include a $\delta$-function singularity at $\epsilon=\mu < 0$ according to
\begin{equation}
	%n_\text{i}(\epsilon)=N_\text{i}\,\theta(1-\epsilon/Q_\text{s})\,\theta(\epsilon)\,.
	n_\text{i}(\epsilon)=\theta(1-\epsilon/Q_\text{s})\,\theta(\epsilon-\mu)+\delta(\epsilon-\mu)\,.
	\label{inix}
\end{equation}
The $\delta$-function singularity in the initial conditions of the NBDE has an analogous role as the singularity that can be added to the Boltzmann equation in order to act as a seed condensate \cite{setk95} since the time evolution of the solutions without singularity does not lead to condensate formation. In Ref.\,\cite{gw20a}, another choice of the initial conditions had been explored, with a thermal distribution in the negative-energy region that also has a singularity at $\epsilon=\mu$. Although the results differ in detail, the overall representation of thermalization is similar to the present results.

\begin{figure}[t!]
	\centering
	\includegraphics[scale=0.42]{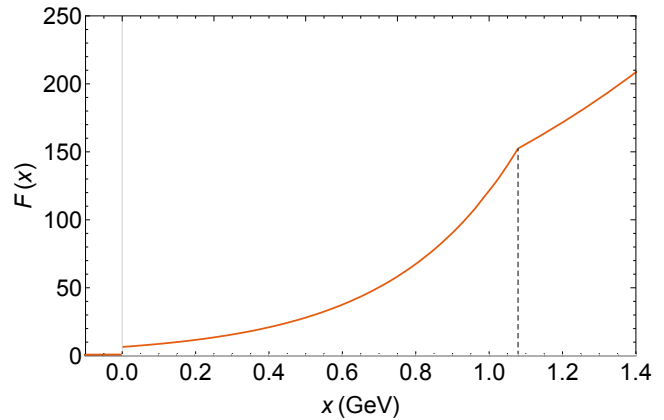}
	\caption{\label{fig2}
		The function $F(x)$ of Eq.\,(\ref{fini}) (solid curve) with $x\rightarrow x+\mu$ as required by the boundary conditions, and a singularity at the origin. $F(x)$ contains the integral over an initial nonequilibrium gluon distribution $n_\text{i}(x)$ according to Eq.\,(\ref{fxmu}). The parameters are given in the text.
		%It recovers the equilibrium distribution with temperature $T$ at $t\simeq 40$ ms.
	}
\end{figure}

The time-dependent partition function with the above initial condition can now be calculated using the bound Green's function Eq.\,(\ref{Greens}), and the function $F(x)$ from Eq.\,(\ref{fini}). The latter contains an indefinite integral over the initial condition Eq.\,(\ref{inix}) that can be carried out to obtain (with $x\rightarrow x+\mu$ in the argument of $F(x)$ as required by the boundary conditions)
\begin{multline}
	F(x)=\exp\left[\frac{-v(x+\mu)}{2D}\right] \\
	\times\exp[-(v/D)
	\,\theta(x)((\mu - Q_s)\theta(\mu -Q_s)\\ + (Q_s - x - \mu)
	\,\theta(x + \mu - Q_s) + x + 1)]\,.
	\label{fxmu}
\end{multline}
%with the auxiliary functions
%\begin{widetext}
%\begin{align}
%\small
%F_1(x)&=\left(\exp(-\mu/T)-\exp\Bigl[\frac{-x-\mu}{T}\Bigr]\right)\theta(Q_\text{s}-\mu-x)+\left(\exp(-\mu/T)-\exp(-Q_\text{s}/T)\right)\theta(x+\mu-Q_\text{s}),\\
%%F_2(x+\mu)&=\exp\Bigl[(-v/D)\theta(x+\mu)((Q_\text{s} - x - \mu)\theta(x + \mu - Q_\text{s}) - Q_\text{s}\theta(-Q_\text{s}) + x + \mu)\Bigr].
%F_2(x)&=\exp\Bigl[(-v/D)\theta(x+\mu)((Q_\text{s} - x - \mu)\theta(x + \mu - Q_\text{s})+ x + \mu)\Bigr].
%\label{tini}
%\end{align}
%\end{widetext}
The function $F(x)$ is plotted in Fig. \ref{fig2}. Due to the singularity in its argument, $F(x)$ has a discontinuity at $x=0$. It is continuous, but not differentiable at $x=Q_\text{s}-\mu$. Both properties are essential to account for the equilibration near the singularity, and in the UV region.
The Green's function of Eq.\,(\ref{Greens}) that includes the IR boundary condition can explicitly be written as
\begin{equation}
	G(\epsilon,x,t)=\exp\left[\frac{-(\epsilon - \mu - x)^2}{4Dt}\right]
	-\exp\left[\frac{-(\epsilon - \mu + x)^2}{4Dt}\right].
	\label{greens}
\end{equation}
With $F(x)$ and $G(\epsilon,x,t)$, the partition function $\mathcal{Z}(\epsilon,t)$ of Eq.\,(\ref{eq:newformulaforZ}) and its derivative $\partial\mathcal{Z}/\partial\epsilon$ can now be calculated, as well as the occupation-number distribution $n(\epsilon,t)$ from Eq.\,(\ref{eq:Nformula}). The full calculation may in principle be carried out analytically. In the case of initial conditions that are appropriate for evaporative cooling of atomic Bose gases at very low energy, we have performed such an exact calculation including the boundary conditions at the singularity in Ref.\,\cite{rgw20}. Here I compute the partition function and its derivative using the \texttt{NIntegrate} and \texttt{Derivative} routines of Mathematica.
\subsection{Discussion of the Solutions for Gluons}
The bosonic equilibration time $\tau_\text{eq}$ is taken as $\tau_\text{eq}=4D/(9v^2)\simeq 0.1$ fm/$c$. This expression has been determined in Ref.\,\cite{gw18} for a $\theta$-function initial distribution in the UV.
% provided the singularity at $\epsilon=\mu<0$ is disregarded.
\begin{figure*}[t]
	%\begin{widetext}
	\centering
	\includegraphics[scale=0.42]{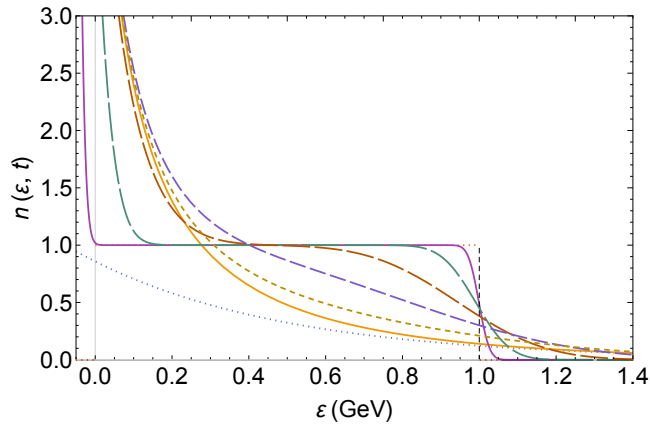}% Here is how to import EPS art
	\caption{\label{fig3}
		Local thermalization of gluons as represented by time-dependent solutions of the nonlinear boson diffusion equation (NBDE) for $\mu<0$. Starting from schematic initial conditions Eq.\,(\ref{inix}) in the cold system at $t=0$ (box distribution with cut at $\epsilon=Q_\text{s}=1$ GeV), a Bose--Einstein equilibrium distribution with temperature $T=513$ MeV (lower solid curve) is approached. Time-dependent single-particle occupation-number distribution functions are shown at $t = 6\times10^{-4}$, $6\times10^{-3}$, $0.04$, $0.12$, and $0.36$ fm/$c$ (decreasing dash lenghts). The lower dotted curve is Boltzmann's distribution.
		%The equilibrium distribution at temperature $T_\text{f}$ is reached within $t\simeq 14$ ms.
	}
	%\end{widetext}
\end{figure*}
The approach to equilibrium provided by the solutions of the NBDE for the gluon distribution functions is shown in Fig.\,\ref{fig3} at $t = 6\times10^{-5}$, $6\times10^{-4}$, $6\times10^{-3}$, $4\times10^{-2}$, $0.12$, and $0.36$ fm/$c$, with decreasing dash lenghts.
%Thermalization occurs much faster than in the linear RTA case.
The steep cutoff in the UV at $\epsilon = Q_\text{s}$ is smeared out at short times -- this was the case already in the free solution without boundary conditions \cite{gw18}. The diffusion coefficient is $D = 1.17$ GeV$^2c$/fm, the drift coefficient $v=-2.28$ GeV$c$/fm.
Correspondingly, the equilibrium temperature in this model calculation is $T=-D/v\simeq513$ MeV $=0.513$ GeV, as expected for the initial central temperature in a Pb-Pb collision at the LHC energy of $\sqrt{s_{NN}} = 5$ TeV \cite{hnw17}. Solutions for related, but different initial conditions (a thermal distribution in the energy region $\mu \le \epsilon \le 0$) have been discussed in Ref.\,\cite{gw20a}.

%The bosonic local equilibration time in this calculation is the same as the one taken for the linear RTA result in Fig.\,\ref{fig1}, $\tau_\text{eq} = 4D/(9v^2)\simeq 0.1$ fm/$c$. This %is the time constant for reaching thermal equilibrium in the UV tail of the distribution function. It may take somewhat longer to attain equilibrium in the IR region, as is the case in the %present model calculation. For given local temperature $T$ and equilibration time $\tau_\text{eq}$, the transport coefficients in the NBDE are obtained as
%\begin{equation}
%D = \frac{4}{9\,\tau_{\text{eq}}} \, T^2 \, , \;\;\;\;\;\;\;\;\; v = -\frac{4}{9\,\tau_{\text{eq}}} \, T \,,\label{tr2}
%\end{equation}
%which for $\tau_{\text{eq}}=0.1$ fm/$c$ corresponds to the values chosen above. The value of the gluonic chemical potential $\mu=-0.36$ GeV has been adapted such that for $T=513$ MeV the initial condition at
%$\epsilon=0$ becomes $n_\text{i}(0)=1$.
%From Figure\,\ref{fig3} it is obvious that the inital nonequilibrium distribution $n_\text{i}(\epsilon)$ gradually approaches the local thermal equilibrium $n_\text{eq}(\epsilon)$ at %$T=513$ MeV
%through the solutions of the NBDE. As discussed already in Ref. \cite{gw18}, these solutions are expected to provide a more realistic description of the thermalization than the %relaxation time approximation (RTA), which enforces a linear approach from $n_\text{i}$ to $n_\text{eq}$, and cannot smoothen the initial discontinuities at the UV cutoff.
The assumption of a constant negative chemical potential $\mu<0$ used in this work is, of course, an idealization that facilitates analytical solutions of the nonlinear problem.
Here, the value of $\mu$ is calculated from particle-number conservation
\begin{equation}
	%n_\text{i}(\epsilon)=N_\text{i}\,\theta(1-\epsilon/Q_\text{s})\,\theta(\epsilon)\,.
	N_\text{i}=N_\text{f}=\int_0^\infty n_\text{eq}(\epsilon)\,g(\epsilon)\,d\epsilon
	\label{ntot}
\end{equation}
with the density of states $g(\epsilon)$. For constant density of states and the above parameter values ($T=513$ MeV), the result is $\mu=-0.08$ GeV. For the more realistic density of states $g(\epsilon)\propto \epsilon^2$ that is valid for a zero-mass relativistic system of gluons, the particle number is then also approximately conserved.

In general, particle-number conservation is strictly fulfilled e.\,g.~for atomic Bose gases at much lower energy, but not for gluons in high-energy collisions.
Driven by particle-number conservation, cold bosonic atoms can move into the condensed phase, thus diminishing the number of particles in the thermal cloud. The chemical potential in the equilibrium solution of the NBDE then becomes time dependent, as has been discussed in Ref.\,\cite{rgw20}, albeit without a full quantum treatment of the condensed phase. It would become zero only in the limit of an inifinite number of particles in the condensed phase. Instead, it approaches a small but finite negative value for a finite number of particles.

In case of relativistic heavy-ion collisions, however, gluons can be created and destroyed. It is therefore unlikely that a condensed phase is actually formed, as had been proposed in model investigations where only soft elastic, number-conserving gluon collisions were considered \cite{jpb12}. Gluon condensate formation in relativistic collisions is essentially prevented by number-changing inelastic processes that correspond to splitting and merging of gluons, although a transient condensate formation is still being debated \cite{blmt17}.

Hence, since inelastic collisions cannot be neglected, the gluon equilibrium distribution is expected to have a nearly vanishing, but still slightly negative, chemical potential which should be approached by the time-dependent solutions of the NBDE. It would therefore be of interest to repeat the present calculation for a time-dependent chemical potential, with $\mu(t)\rightarrow0$ for $t\rightarrow\infty$, as was done in Ref.\,\cite{rgw20} for the case of cold atoms. This requires, however, numerical work that goes substantially beyond the present analytic approach.
% where I consider an analytic calculation for the limiting case $\mu=\text{const}<0$,
%replacing the initial condition of Eq.\,(\ref{inix}) with a box distribution plus a delta-function at $\epsilon=\mu<0$, and boundary conditions at the singularity. For this special limiting %case
%with $n_\text{eq}(\epsilon=\mu\rightarrow 0)=\infty$, it is possible to obtain closed-form solutions of the NBDE. Solutions for related, but different initial conditions (a thermal %distribution in the energy region $\mu \le \epsilon \le 0$) have been discussed in Ref.\,\cite{gw20a}.
\subsection{Discussion of the Solutions for Quarks}
\begin{figure}[t!]
	\centering
	\includegraphics[scale=0.42]{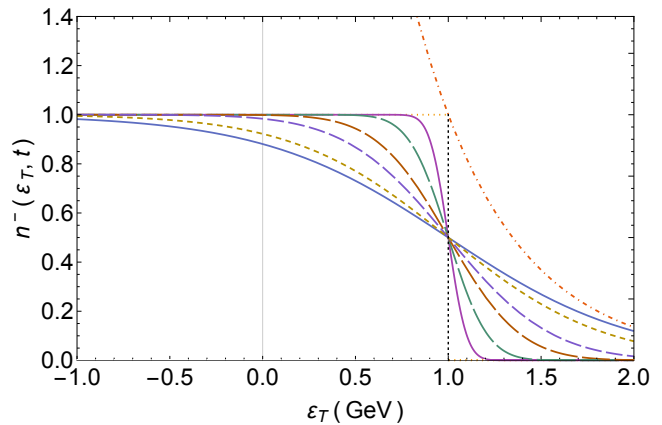}
	\caption{\label{fig4}
		Occupation-number distribution $n^-(\epsilon_\text{\,T},t)$ of a relativistic fermion system (valence quarks) as function of the transverse energy $\epsilon_\text{\,T}$
		%(valence quarks in the local rest frame of a fragmentation source)
		including antiparticle production ($\epsilon_\text{\,T}<0)$. It is evaluated analytically from Eq.\,(\ref{bose}) at different times for the initial distribution $n^-_\text{i}(\epsilon_\text{\,T}) = \theta\,(1 - \epsilon_\text{\,T}/\mu)$. The parameters are $\mu= 1 \, \text{GeV}$, $T = -D/v = 500 \, \text{MeV}$, $\tau^-_\text{eq}\,=4D/v^2\,= 0.9$ fm/$c$. The distribution $n^-(\epsilon_\text{\,T},t)$ is displayed at
		%$t/\tau_{\text{eq}}=$ 0, 0.003, 0.015, 0.06, 0.15,0.6, $\infty$ (ordered by decreasing dash length).
		%$tc/$fm =
		$t = 0.003$, $0.015$, $0.06$, $0.15$, and $0.6$ fm/$c$ (ordered by decreasing dash length towards the solid equilibrium distribution). The upper dot-dashed curve is Boltzmann's distribution.
		The negative-energy region corresponds to antiparticle production. From Ref.\,\cite{gw19}.
	}
\end{figure}

The local thermalization of valence quarks towards the Fermi--Dirac equilibrium distribution is easier to calculate in the nonlinear model because no singularity occurs for fermions.  Hence, the free solutions can be used as was already discussed in Ref.\,\cite{gw18} and in more detail in Ref.\,\cite{bgw19}. Hadron production in the early thermalization phase was implicitly considered, because the negative-energy region corresponds to particle-antiparticle production. A typical result for the fermionic time-dependent occupation-number distribution as function of the transverse energy taken from Ref.\,\cite{gw19} is shown in Fig.\,\ref{fig4}. Local thermalization for quarks occurs more slowly as for gluons due to Pauli's principle, and also because of the larger color factor for gluons.
\begin{figure}
	\centering
	\includegraphics[scale=0.6]{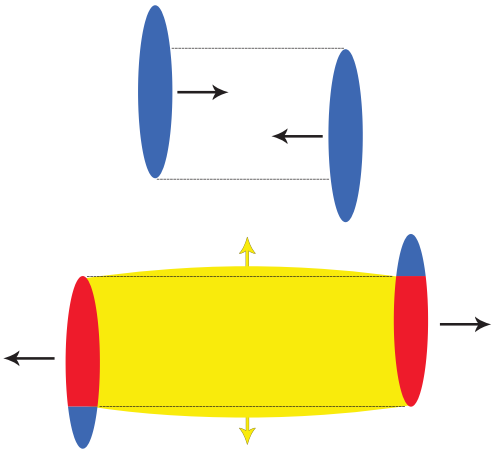}
	\caption{\label{fig5}
		Schematic representation of the three-source model for relativistic heavy-ion collisions at RHIC and LHC energies in the center-of-mass system: Following the collision and slowing down (\textit{stopping}) of the two Lorentz-contracted slabs (blue), the fireball region (center, yellow) expands anisotropically in longitudinal and transverse direction. At midrapidity, it represents the main source of particle production. The two fragmentation sources (red) contribute to particle production, albeit mostly in the forward and backward rapidity regions. In stopping, these are the only sources. From Ref.\,\cite{kgw19}.
	}
\end{figure}
%\section{Conclusion}
%Analytic solutions of the nonlinear boson diffusion equation have been explored for the thermalization of gluons in relativistic hadronic collisions. The solutions take account of a %singularity in the initial conditions at $\epsilon=\mu<0$ with fixed chemical potential $\mu$, and the boundary conditions at the singularity.

%Different from earlier results that were calculated with the free Green's function and converged to the Bose--Einstein equilibrium solution only in the UV, these solutions converge %towards Bose--Einstein also in the IR and hence, properly describe thermalization in a finite gluon system.

The bounded solutions of the NBDE and the corresponding nonlinear fermion diffusion equation are, in particular, tailored to local thermalization processes that occur in relativistic heavy-ion collisions at energies reached at RHIC and LHC. In the present example, they are applied to the local equilibration of quarks and gluons in central Pb-Pb collisions at a center-of-mass energy of 5 TeV per nucleon pair, leading to rapid thermalization with a local temperature of $T\simeq500$ MeV.
Since the thermalization occurs very fast -- before anisotropic expansion fully sets in --, the analytic solution of the problem in 1+1 dimensions appears permissible. The hot system will subsequently expand anisotropically and cool rapidly, as is often modeled successfully by relativistic hydrodynamics \cite{hesne13}, until hadronization is reached at $T\simeq160$ MeV.
%In conclusion, my schematic model based on the NBDE accounts for the fast nonlinear approach to local thermal equilibrium from an initial nonequilibrium gluon distribution at the %start of the collision.
%It avoids the discontinuities that are inherent in the well-established relaxation time approximation, which enforces a linear approach to equilibrium.
Further refinements of the thermalization model such as time-dependent transport coefficients are conceivable, but are unlikely to allow for analytic solutions. A microscopic calculation of the transport coefficients with an investigation of their dependencies on energy and time would be very valuable. Extensions of the NBDE itself to higher dimensions in order to account for possible anisotropies should also be investigated.
\section{Stopping: Net-Proton Distributions}
The incoming baryons with energies available at SPS, RHIC or LHC are being \textit{stopped} on an equally short time scale as the local equilibration occurs: In the course of the collision shown schematically in Fig.\,\ref{fig5}, the system is being slowed down, essentially through collisions of the incoming valence quarks with soft gluons in the respective other nucleus. Various models to account for this process and its energy dependence have been developed, in particular in Refs.\,\cite{mtw09,mtwc09} and related works which are relying on the appropriate parton distribution functions (PDFs).
They yield agreement with the available net-proton (proton minus antiproton) stopping data. Different from the nonequilibrium-statistical approach to initial thermalization, such models do not consider a time dependence. However, by using the rapidity distribution calculated from the PDFs, and the initial valence-quark distribution, one can, in addition, account for the time development from the initial to the final distribution with an appropriate fluctuation--dissipation relation.

The fragmentation peaks in stopping occur mainly due to the interaction of valence quarks with soft gluons in the respective other nucleus.
For net protons (protons minus antiprotons), their positions in rapidity space $y=0.5\ln[(E_{||}+p_{||})/(E_{||}-p_{||})]$  with $m_\text{p}=m_\text{\textoverline{p}}$ can be obtained from Ref.\,\cite{mtw09}, and references therein, as
\begin{equation}
	\frac{\mathrm{d}N_\mathrm{p-\bar{p}}}{\mathrm{d}y}=\frac{C}{(2\pi)^2}\int\frac{\mathrm{d}^2p_\mathrm{T}}{p_\mathrm{T}^2}x_1q_v(x_1,p_\mathrm{T})f_g(x_2,p_\mathrm{T})\,.
	\label{frag}
\end{equation}
This expression accounts for the peak in the forward region, and there is a corresponding one with $y\rightarrow -y$ for the backward peak. The longitudinal momentum fraction of the valence quark $v$ that experiences stopping is $x_1 = p_\mathrm{T}/\sqrt{s}\exp(y)$, the one for the soft gluon $g$ in the target is $x_2 = p_\mathrm{T}/\sqrt{s}\exp(-y)$. The distribution function of the valence quarks is $q_v(x_1,p_\mathrm{T})$, the one of the gluons is $f_g(x_2,p_\mathrm{T})$. The latter represents the Fourier transform of the forward dipole scattering amplitude $N(x_2,r_\mathrm{T})$ for a quark dipole of transverse size $r_\mathrm{T}$ \cite{mtwc09}. To account for the correct normalization in net-proton or net-baryon distributions, the  normalization constant $C$  is adjusted such that the integral of Eq.\,(\ref{frag}) agrees with the total number of participant protons or baryons.

Fragmentation peaks are then found to occur at $y=\pm y_{\text{peak}}$ in rapidity space. At suffiently high energy -- in particular, at LHC energies -- these positions become sensitive to the gluon saturation scale
\begin{figure}
	\centering
	\includegraphics[scale=0.62]{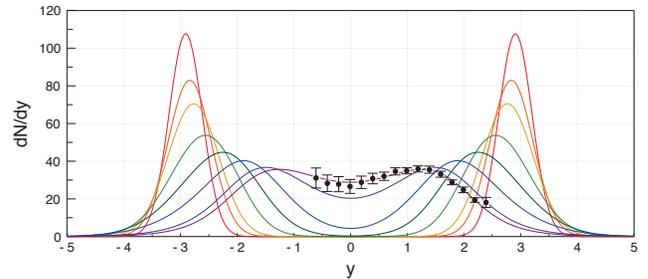}
	\caption{\label{fig5a}
		Rapidity distributions of net protons in central Pb-Pb collisions at SPS energies of $\sqrt{s_{NN}}$ = 17.3 GeV compared with NA49 data \cite{app99}. Red curves are the initial distributions broadened by the Fermi momentum,
		% red-dashed curves the asymptotic distributions for $t\rightarrow \infty$, 
the final distribution is from a QCD-inspired model \cite{mtw09} with the saturation-scale exponent $\lambda=0.2$ and $Q_0^2=0.09$ GeV$^2$ (see text). It agrees with the data, and corresponds to the distribution at the interaction time $t=\tau_\text{int}$ in a time-dependent formulation. The six intermediate solid curves at $t/\tau_\text{int} = 0.01, 0.02, 0.05, 0.1, 0.2$ and $0.5$ account for the time dependence in a nonequilibrium-statistical relativistic theory. From Hoelck and Wolschin \cite{hgw20}.
	}
\end{figure}
%\begin{figure}[tph]
%	\centering
%	\includegraphics[width=6cm]{fig3}
%	\caption{\label{fig3}(Color online)
%		Evidence for fragmentation sources: Rapidity distributions of net protons in central Pb-Pb collisions at SPS energies of $\sqrt{s_{NN}}$ = 17.3 GeV (top frame) compared with NA49 data \cite{app99}. Solid curves correspond to a gluon saturation momentum $Q_\mathrm{s} = 0.9$ GeV/$c$ at $x = 0.01$, dashed curves to $Q_\mathrm{s} = 1.8$ GeV/$c$.
%		At RHIC energies of 62.4 GeV (middle frame) and 200 GeV (bottom frame) for central Au-Au, theoretical results are compared with BRAHMS net proton data~\cite{bea04}. The fragmentation peaks move apart in rapidity space with increasing energy. Arrows indicate the beam rapidities. From Mehtar-Tani and Wolschin \cite{mtw09,mtw15}.
%	}
%\end{figure}
%The fragmentation peak positions $y_{\text{peak}}$ in rapidity space are at suffiently high energy -- in particular, at LHC energies -- indicators for the gluon saturation scale
\begin{equation}
	Q_\mathrm{s}^2 = A^{1/3}Q_0^2x^{-\lambda}\,.
\end{equation}
Here, $A$ is the mass number, $Q_0$ the momentum scale, $x<1$ the momentum fraction carried by the gluon, and $\lambda$ the saturation-scale exponent.

values of the gluon saturation scale and found to agree with net-proton data from SPS and RHIC. We are presently incorporating these results into a time-dependent nonequilibrium-statistical relativistic theory \cite{hgw20}.  A typical outcome of our approach for central Pb-Pb at SPS energies of $\sqrt{s_{NN}}=17.3$ GeV is shown in Figure\,\ref{fig5a} where it is compared with NA49 data \cite{app99} in rapidity space. The initial distributions are shown as red curves, with the initial broadening due to the Fermi motion. The final distribution agrees with the one from our QCD-inspired model \cite{mtw09} with a saturation-scale exponent $\lambda=0.2$ and $Q_0^2=0.09$ GeV$^2$, translating into a gluon saturation momentum of $Q_\mathrm{s}\simeq 0.55$ GeV/$c$ at $x=10^{-4}$. At the interaction time $t=\tau_\text{int}$, the time-dependent distribution agrees with the data.
The six intermediate solid curves at $t/\tau_\text{int} = 0.01, 0.02, 0.05, 0.1, 0.2$ and $0.5$ show the time dependence in our nonequilibrium-statistical relativistic theory \cite{hgw20}.

A larger gluon saturation momentum $Q_\mathrm{s}$ was found to produce more stopping, as does a larger mass number $A$ \cite{mtw09}.
In the context of an investigation of particle production, the agreement between the calculated stopping distributions and the data will be taken as evidence for the importance of fragmentation contributions also in charged-hadron production.

In Ref.\,\cite{mtw11}, Mehtar-Tani and I found that the fragmentation peak positions in stopping depend in a large c.m.~energy range 6.3 GeV $\leq \sqrt{s_{NN}} \leq 200$ GeV linearly on the beam rapidity $y_{\text{beam}}$ and the saturation-scale exponent $\lambda$ according to 
\begin{equation}
	y_{\text{peak}}=\frac{1}{1+\lambda}(y_{\text{beam}} - \ln A^{1/6}) + \mathrm{const}\,.
\end{equation}
At the current LHC energy of 5.02 TeV Pb-Pb corresponding to $y_{\text{beam}} = \pm \ln({\sqrt{s_{NN}}/m_\mathrm{p}}) = \pm~8.586$ and with a gluon saturation-scale exponent $\lambda \sim 0.2$ one therefore expects $y_{\text{peak}}\simeq \pm~6$.
Due to the lack of a suitable forward spectrometer at LHC, the rapidity region of the peaks will thus not be accessible for identified protons in the coming years at LHC energies. Nevertheless, the partonic processes that mediate stopping also contribute to hadron production at LHC energies and hence, one expects fragmentation events in particle production. Experience with stopping versus hadron production at SPS and RHIC energies \cite{gw13,gw16} has shown that the fragmentation peaks in particle production occur consistently at somewhat smaller absolute rapidities than the ones in stopping.

In net-baryon (proton) distributions charged baryons produced from the gluonic source cancel out because particles and antiparticles are generated in equal amounts. In charged-hadron production, however, this is not the case. Instead, three sources contribute provided the energy is sufficiently high, $\sqrt{s_{NN}} > 20$ GeV.  It was found in Ref.\,\cite{gw16}     that the dependence of their particle content on c.m.~energy differs: The fragmentation sources contain $N_\mathrm{ch}^{qg} \propto \ln(s_{NN}/s_0)$ charged hadrons, whereas the midrapidity-centered source that arises essentially from the interaction of low-$x$ gluons contains $N_\mathrm{ch}^{gg} \propto \ln^3(s_{NN}/s_0)$ charged hadrons. Due to the strong $\propto \ln^3$ dependence, it  becomes more important than the fragmentation sources at LHC energies \cite{gw15}.

With the three sources, the total rapidity distribution for produced charged hadrons becomes
%Since the fragmentation distributions must exist in charged-hadron production because they can be measured separately in net-proton data, and the gluonic distribution is known to %be present in particle production, with particles and antiparticles produced in equal amounts, the total rapidity distribution for produced charged hadrons becomes
\begin{multline}
	\frac{\mathrm{d}N^{\mathrm{tot}}_{\mathrm{ch}}(y,t=\tau_{\mathrm{int}})}{\mathrm{d}y}=
	N_{\mathrm{ch}}^{{qg,1}}R_{1}(y,\tau_{\mathrm{int}}) \\
	+N_{\mathrm{ch}}^{{gq,2}}R_{2}(y,\tau_{\mathrm{int}})+N_{\mathrm{ch}}^{gg}R_{gg}(y,\tau_{\mathrm{int}})\,.
	\label{normloc1}
\end{multline}
The fragmentation distributions $R_{1,2}(y,t)$ and gluonic distributions $R_{gg}(y,t)$ can be calculated in a time-dependent phenomenological model such as the relativistic diffusion model \cite{gw13}, or in microscopic theories.
The strong interaction ceases to act at the interaction time ($\equiv$ freezeout-time) $t = \tau_\mathrm{int}$ and theoretical distributions may be compared to data in a $\chi^2$-optimization.

A transparent phenomenological model to calculate and predict the distribution functions of produced particles is the
relativistic diffusion model \cite{wolschin99,gw13}. In the RDM, the initial distribution functions are evolved up to $\tau_{\mathrm{int}}/\tau_y$ with the rapidity relaxation time $\tau_y$ using the analytical moments equations. The mean values $\langle y_{1,2} \rangle$ of the fragmentation distributions that are related to $\tau_{\mathrm{int}}/\tau_y$ can be determined
%in $\chi^2$-minimizations with respect to
from the data. The details of the model calculations are given in the corresponding Refs.\,\cite{gw13,gw15,gw16}.
%The absolute value of $\tau_{\mathrm{int}}$ does not appear in this calculation because it would require a theory for $\tau_y$, which is not available to date.
%(The value 5-8 fm/c is as determined e.\,g.~in HBT experiments
%but not needed explicitly here).
\section{Charged-Hadron Production} 
\subsection{Transverse-momentum distributions}
As discussed frequently in experimental papers \cite{qm19} and also in Ref.\,\cite{gw16}, the transverse-momentum distributions of produced charged hadrons in relativistic heavy-ion collisions
show an exponential behaviour in the thermal regime as accounted for by the Maxwell--J\"uttner distribution \cite{juett11} ($c\equiv 1$) 
\begin{equation}
f(p_\mathrm{T})=\frac{1}{4\pi m^2 T K_2(m/T)}\exp{\left[-\frac{\gamma(p_\mathrm{T})m}{T}\right]}
\label{juett}
\end{equation}
with the modified Bessel function of the second kind $K_2(m/T)$, the Lorentz-factor
\begin{equation}
\gamma(p_\mathrm{T})=\sqrt{1+(p_\mathrm{T}/m)^2},
\end{equation}
freeze-out
temperature $T$ and 
hadron mass $m$.
Beyond $p_\text{T} \simeq 4$ GeV/$c$, however, a transition 
 to a power-law  (straight lines in a log-log plot) occurs. It is attributed mostly to the recombination of soft partons, and fragmentation of hard partons. 
In addition to detailed theoretical approaches, this transition can be modelled phenomenologically using distribution functions of the QCD-inspired form (see references in Wilk and Wong \cite{wowi13}) %\cite{wowi13}
used by Hagedorn \cite{hag83}
for high-energy pp and p\textoverline{p} collisions 
 \begin{equation}
E\frac{{\text{d}^3 \sigma}}{\text{d}p^3} = C\, (1 + p_\mathrm{T}/p_0)^{-n}
\label{hag}
\end{equation}
with a normalization constant $C$ and parameters $p_0, n$.
This expression describes the transition from exponential 
%($\propto \exp(-p_\mathrm{T}/T)$
for $p_\mathrm{T}\rightarrow 0$ 
as in the J\"uttner distribution Eq.\,(\ref{juett}) (with $p_0=nT$ and $p_\text{T} \rightarrow m_\text{T}$), to power-law behaviour
($\propto  (p_\mathrm{T}/p_0)^{-n}$ for $p_\mathrm{T}\rightarrow \infty$) . 

\begin{figure}[t]
\centering
\includegraphics[scale=0.3]{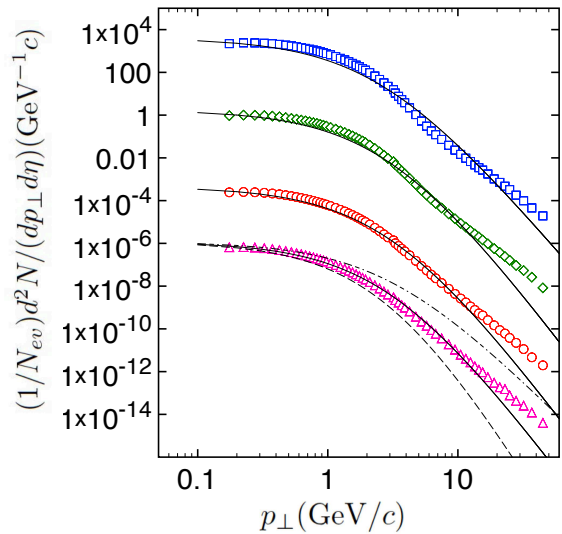}
\caption{ Transverse momentum distributions of produced charged hadrons in $\sqrt{s_{NN}}=5.02$ TeV Pb-Pb collisions calculated from Hagedorn's formula (see text) compared with ALICE data \cite{alice18} for 0--5\%, 20-30\% ($\times 10^{-3}$), 50--60\% ($\times10^{-6}$) and 70--80\% ($\times10^{-8}$) centralities (top to bottom). The dot-dashed curve has power index $n=6.2$, the dashed curve $n=10.2$, solid curves are for $n=8.2$. Error bars are smaller than the symbol size.}
%  for $q = 1.12$ and $T = 120$ MeV 
%that implicitly account for emission during collective expansion, 
%but not for hard (pQCD) processes at $p_\mathrm{T} \ge 7$ GeV/$c$. 
%Solid curves are for $q = 1.10$, the dashed curve is for $q = 1.12$. Peripheral spectra are scaled for better visibility, see Figure~1 for absolute values.}
% and RDM parameters as in table 1. 
%The solid curve is a $\chi^2$-minimization based on the three-sources RDM with respect to the preliminary %ALICE data from \cite{to11}
%that takes the limiting fragmentation hypothesis into account, see text. Error bars have been symmetrized %using the larger branches. The corresponding 
%RDM parameters are given in Table~\ref{tab1}, and ~2. 
%, with the dashed curve
%arising from gluon-gluon collisions in the fireball, the dash-dotted curve from valence quark-gluon events in the Pb-like region, and the dotted curve correspondingly in the proton-like %region.
\label{fig6} 
\end{figure}

Using Eq.~(\ref{hag}), Fig.~\ref{fig6} shows $p_\mathrm{T}$-distributions of produced charged hadrons at four centralities in 5.02 TeV Pb-Pb compared with ALICE data \cite{alice18} (peripheral spectra are scaled for better visibility; statistical and systematic error bars are smaller than the symbol size). 
%Here the freezeout-temperature is $T \equiv T_\mathrm{F} = 120$ MeV and the average mass is
%$m \equiv \langle m \rangle = 0.22$ GeV/$c^2$, as in Fig.~\ref{fig1}. 
The data are well represented through many orders of magnitude with a power index $n = 8.2$ and $p_0=3$ GeV/$c$ (Fig.~\ref{fig6}), but at high $p_\text{T}$ deviations occur which are
attributed to hard processes that require a pQCD treatment. This corresponds to the occurence of a minimum in the nuclear modification factor for produced charged hadrons as function of $p_\mathrm{T}$ found already at $\sqrt{s_{NN}}=2.76$ TeV \cite{abe13}, and confirmed at 5.02 TeV \cite{alice18}.

There is presently no theoretical derivation for the
value of the power index $n$  that is needed to reproduce the experimental $p_\mathrm{T}$-distributions in relativistic heavy-ion collisions. It is therefore not obvious 
from the present analysis which fraction of low-$p_\mathrm{T}$ particles could be due to nonequilibrium processes 
%and/or hard pQCD events 
that differ from thermal emission out of a single expanding fireball. In particular, one can not distinguish particles emitted from the
fireball and those arising from the fragmentation sources at low $p_\mathrm{T}$. Hence, the analysis of transverse momentum distributions in terms of Eq.\,(\ref{hag}) 
is presently only suitable to distinguish high-$p_\mathrm{T}$ hard events from the bulk of (thermal and nonequilibrium) charged-hadron emission.
\subsection{Pseudorapidity distributions}
The distinction of particles emitted from the fireball and those from the fragmentation sources is more transparent
in rapidity or pseudorapidity distributions of produced charged hadrons. The existence of the fragmentation sources is 
evident from the measurements of stopping in heavy-ion collisions as discussed in Section 3.
Pseudorapidity distributions $\mathrm{d}N_\text{ch}/\mathrm{d}\eta$ with $\eta = - \ln\,[\tan(\theta/2)]$ depend only on the scattering angle $\theta$. 
Particle identification is not needed here and hence, they are much easier to obtain at large $\eta$-values (small scattering angles) compared to rapidity distributions.
%which are defined for each particle species separately -- at large 
%values of $y$.
%$y = 0.5\,\ln[(E+p_{\parallel})/(E-p_{\parallel})]$.
%To assess the significance of the fragmentation sources in particle production at LHC energies, it is therefore better to compare theoretical models
%with pseudorapidity distributions of produced charged hadrons, rather than rapidity distributions of identified particles.

\begin{figure}[t]
\centering
\includegraphics[width=8.6cm]{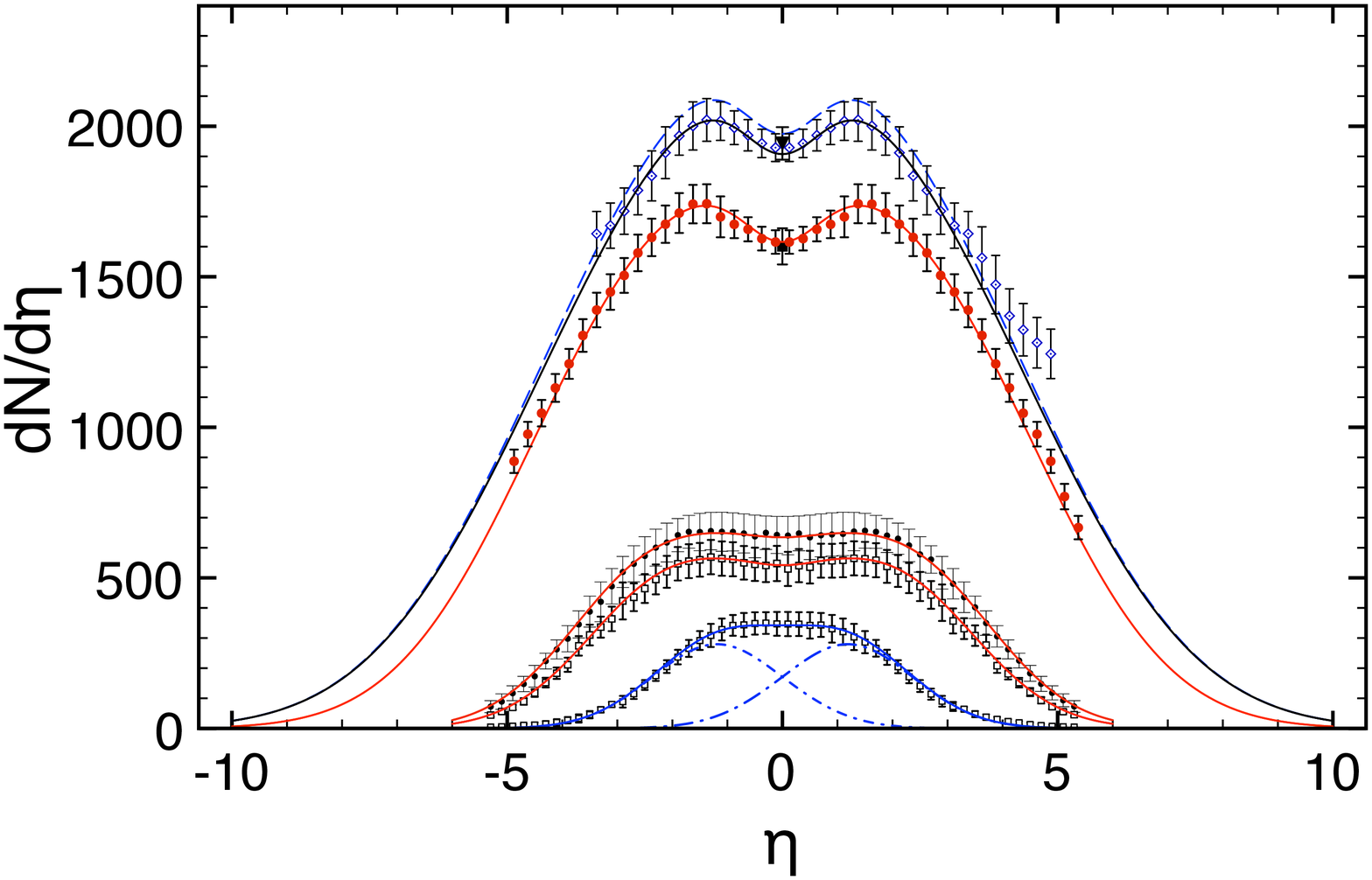}
\caption{ The RDM pseudorapidity distribution functions for charged hadrons in central Au-Au (RHIC) and Pb-Pb (LHC) collisions
at c.m.~energies of 19.6 GeV, 130 GeV, 200 GeV, 2.76 TeV, and 5.02 TeV shown here are optimized in $\chi^2$-fits with respect to the PHOBOS \cite{bb03,alv11} (bottom) and ALICE \cite{abb13,alice17} (top) data, with parameters from Refs.\,\cite{gw13,gw16}. The upper dashed distribution function at 5.02 TeV is a prediction from Ref.\,\cite{gw16} within the relativistic diffusion model. The 5.02 TeV midrapidity data point is from Ref.\,\cite{ad16}, the  5.02 TeV data from Ref.\,\cite{alice17}. The 19.6 GeV distribution has only two sources
(dot-dashed), the other ones have three.}
\label{fig7}
\end{figure}

The pseudorapidity distributions $dN_\mathrm{ch}/d\eta$ for produced charged hadrons in relativistic heavy-ion collisions
emerge from a superposition of the fragmentation sources and a midrapidity source. According to the discussion in Ref.\,\cite{gw15} and at the end of Section\,3,
the particle content of the low-$x$ gluon source rises rapidly according to $N_{{gg} }\propto \ln^3(s_{NN}/s_0)$. 
To convert rapidity distributions $\mathrm{\mathrm{d}N}_\mathrm{ch}/\mathrm{\mathrm{d}y}$ to pseudorapidity distributions $\mathrm{\mathrm{d}N}_\mathrm{ch}/\mathrm{d}\eta$, 
the corresponding Jacobian is required

\begin{equation}
\frac{\mathrm{d}N}{\mathrm{d}\eta}=\frac{\mathrm{d}N}{\mathrm{d}y}\frac{\mathrm{d}y}{\mathrm{d}\eta}=
%\frac{p}{E}\frac{\mathrm{d}N}{\mathrm{d}y}\simeq
J(\eta,  m / p_\mathrm{T})\frac{\mathrm{d}N}{\mathrm{d}y}\,, 
\label{dJeta}
\end{equation}
\begin{equation}
{J(\eta,m /p_\mathrm{T})=\cosh({\eta})\cdot }
[1+(m/ p_\mathrm{T})^{2}
+\sinh^{2}(\eta)]^{-1/2}\,.
\label{jac}
\end{equation}
The hadron mass is $m$ and the transverse momentum $p_\mathrm{T}$. Since the transformation depends on the squared ratio $(m$/$p_\text{T})^2$ of mass and transverse momentum of the produced particles, its effect increases with the mass of the particles and is most pronounced at small transverse momenta.
% In principle, one has to consider the full $p_\text{T}$-distributions, which are, however, not available for all particle species that are included in the pseudorapidity measurements. 
%A rough estimate is obtained replacing $m$ by the mean mass $\langle m\rangle$, and using the mean transverse momentum $\langle p_\text{T} \rangle$ that is taken from the available %transverse momentum distributions, such as for pions.  
In Ref.\,\cite{rgw12}, we have determined the Jacobian $\mathrm{J}_0$ at $\eta=y=0$ in central 2.76 TeV Pb-Pb collisions for identified $\pi^-, K^-$, and antiprotons from the experimental values $\frac{\mathrm{d}N}{\mathrm{d}\eta}|_\text{exp}$ and $\frac{\mathrm{d}N}{\mathrm{d}y}|_\text{exp}$ as $\mathrm{J}_0=0.856$. We can solve Eq.\,(\ref{jac}) for $p_\text{T}\equiv \langle p_\text{T}^{\text{eff}}\rangle$ to obtain
\begin{equation}
\langle p_\text{T}^{\text{eff}}\rangle=\frac{\langle m \rangle\, \mathrm{J}_0}{\sqrt{1-\mathrm{J}_0^2}}\,.
\label{pteff}
\end{equation}
The mean mass $\langle m \rangle$  can be calculated from the abundancies of pions, kaons, and antiprotons. Using $\mathrm{J}_0$, 
the Jacobian can be written independently from the values of $\langle m \rangle$ and $\langle p_\text{T}^{\text{eff}}\rangle$ as
\begin{equation}
\mathrm{J} \left(\eta, \mathrm{J}_0\right) = 
	\frac{\cosh(\eta) }{\sqrt{ 1 +\frac{1-\mathrm{J}_0^2}{\mathrm{J}_0^2}+ \sinh^2(\eta) }}\,.
	\label{jac1}
\end{equation}
The result for central 2.76 TeV Pb-Pb collisions was found in Ref.\,\cite{rgw12} to be $\mathrm{J}(\eta)=\cosh(\eta)[{1.365+\sinh^2(\eta)}]^{-1/2}$.
% as $\langle m \rangle\simeq \SI{218}{\MeV}$, resulting in $\langle p_\text{T}^{\text{eff}}\rangle \simeq \SI{0.36}{\GeV}$. This is actually smaller than %the mean momentum deduced from the transverse momentum distribution of pions only ($\langle p_\text{T}^{\pi}\rangle \simeq \SI{0.51}{\GeV}$).
The Jacobian affects the pseudorapidity distributions mostly near midrapidity, where it generates a dip as is obvious from Figure\,\ref{fig7}: A prediction in the RDM with linear drift from Ref.\,\cite{gw16} is compared with ALICE data for central Pb-Pb at 5.02 TeV \cite{alice17} (dashed upper curve), and a $\chi^2$-optimization within the five parameter RDM is carried out, upper solid curve. It differs only slightly from the prediction.

The nonequilibrium evolution of all three partial distribution functions $R_k(y,t)$ $(k=1,2,gg)$ towards the thermodynamic equilibrium  
(Maxwell--J{\"u}ttner) distribution
for $t\rightarrow \infty$ 
\begin{multline}
\label{equ}
  E \frac{\text{d}^3N}{\text{d}p^3} \Bigr|_\text{eq}\propto E \exp\left(-E/T\right)\\
  = m_\text{T} \cosh\left(y\right) \exp\left(-m_\text{T} \cosh(y) / T\right)
\end{multline}
is  accounted for in the Relativistic Diffusion Model \cite{wolschin99,biya02,gw13,gw16}  through solutions of the Fokker-Planck equation 
%\EQ{fpe}
\begin{multline}
  \pd{}{t}R_k(y,t) =\\
   -\pd{}{y}\left[J_k(y,t)R_k(y,t)\right] + \pd[2]{}{y}\left[D_k(y,t)R_k(y,t)\right]\,.
\label{fpe}
\end{multline}
The drift functions $J_k(y,t)$ and diffusion functions $D_k(y,t)$ depend on rapidity and time. However, if the diffusion coefficients are taken as constants $D_k$,
and the drift functions assumed to be linearly dependent on the rapidity variable $y$, 
%as $J(y,t)=\alpha y$, 
the FPE acquires the Ornstein-Uhlenbeck form \cite{uo} which can be solved analytically in rapidity space \cite{wolschin99}. In this case it is easy to show that for $t\rightarrow \infty$ all three subdistributions approach a single Gaussian in $y$ space which is centered at midrapidity $y=0$ for symmetric systems, or at the appropriate equilibrium value $y=y_\text{eq}$ for asymmetric systems \cite{gw13,gw15,gw16}. For stopping, only the two fragmentation distributions contribute,  approaching the thermal equilibrium distribution for $t\rightarrow \infty$, as discussed in the previous section. If the system reaches a stationary distribution that differs from the thermal one as discussed in Section\,3 for the case of stopping calculated in a QCD-based model, the underlying fluctuation-dissipation relation becomes more complicated \cite{hgw20}.

%Stopping (or more precisely, slowing down) of the Lorentz-contracted, highly transparent nuclei occurs before the QGP-medium with quarks and gluons in the fireball is fully formed. %Hence, there exists no medium or heat bath that could act as a solvent providing friction and noise due to thermal fluctuations, as is the case in the diffusion model for brownian %motion, or for heavy quarks in a QGP. Instead, the incident baryons loose their momentum (rapidity) without any globally static medium, but through random partonic two-body %collisions between valence quarks and low-$x$ gluons in the respective other nucleus. These provide the fluctuating environment necessary for the formulation of a Langevin %equation, or equivalenty, the corresponding Fokker-Planck equation for the relativistic system.
To derive the above FPE (\ref{fpe}) in the context of relativistic heavy-ion collisions, one can make use of a theory for a special class of non-markovian processes in spacetime discussed in Refs.\,\cite{denisov09,dunkel09}, which are equivalent to relativistic Markov processes in phase space (RMPP). These processes give rise to a generalised FPE which is suitable for describing relativistic diffusive particle dynamics. Our Eq.\,(\ref{fpe}) is conceptually a special case of such a RMPP formalism for charged-hadron production in rapidity space. An application of RMPP to baryon stopping will be shown in Ref.\,\cite{hgw20}.

The $t\rightarrow \infty$ limit of the FPE solution for constant diffusion and linear drift is found to deviate slightly from the Maxwell--J{\"u}ttner distribution. The discrepancies are small and become visible only for sufficiently large times. To ensure that the asymptotic solution yields the Maxwell--J{\"u}ttner distribution Eq.\,(\ref{equ}), a RDM with the sinh-drift is required

\begin{equation}
J_k(y,t)=-A_k \sinh(y)\,,
\end{equation}
as was discussed in Refs.\,\cite{lavagno02,wolschin18}.
The corresponding fluctuation-dissipation relation (FDR) that connects drift and diffusion becomes \cite{wolschin18}
\begin{equation}
A_k= m_\text{T} D_k/T\,.
\label{fdt}
\end{equation}
If the asymptotic distribution is not Maxwell--J{\"u}ttner, but -- as in the case of stopping, where it may be provided by a QCD-based distribution from a calculation as 
performed in Ref.\,\cite{mtw09} --, a different form of the fluctuation-dissipation relation will result, as will be discussed in Ref.\,\cite{hgw20}.

The strength of the drift force in the fragmentation sources $k = 1,2$ depends on the distance in $y$-space from the beam rapidity, which enters through the initial conditions.
With Eq.\,(\ref{equ}), the rapidity distribution at thermal equilibrium can then be derived \cite{wolschin17} as
\begin{multline}
\frac{\text{d}N_\text{eq}}{\text{d}y}=C \left( m_\text{T}^2 T + \frac{2 m_\text{T} T^2}{\cosh y} + \frac{2 T^3}{\cosh^2 y} \right)\\
\times \exp\left({-\frac{m_\text{T} \cosh y}{T}}\right),
\label{eqfdt}
\end{multline}
where $C$ is proportional to the overall number of produced charged hadrons  $N^\text{tot}_\text{ch}$, or -- in case of stopping --  to the number of net baryons (protons) in the where $C$ is proportional to the overall number of produced charged hadrons  $N^\text{tot}_\text{ch}$, or -- in case of stopping --  to the number of net baryons (protons) in the respective centrality bin. Since the actual distribution functions remain far from thermal equilibrium, the total particle number is evaluated based on the nonequilibrium solutions of the FPE, which are adjusted to the data in $\chi^2$-optimizations. 

In particular,
one can determine the drift amplitudes $A_k$ from the position of the fragmentation peaks as inferred from the data, and then calculate theoretical diffusion coefficients as $D_k=A_k T/m_\text{T}$. Since the fireball and both fragmentation sources also expand collectively, the actual distribution functions will, however, be broader than what is obtained from Eq.\,(\ref{fdt}). To account for this broadening through collective expansion, we use
% $D^\text{exp}_k$ 
diffusion coefficients (or widths of the partial distributions) that are adapted to the data in both stopping and particle production. From the integral of the overall distribution function
the total particle number can be obtained. In case of the FPE with sinh-drift, the FPE must be solved numerically
 as described in Ref.\,\cite{wolschin18}. 

Results for central collisions of symmetric systems in the three-source RDM with linear drift are summarized in Figure \ref{fig7}. 
The dependence of the pseudorapidity distributions on c.m.~energy in central Au-Au collisions at 19.6 GeV, 130 GeV, and 200 GeV RHIC energies as well as 
in central Pb-Pb at 2.76 TeV and 5.02 TeV LHC energies are displayed. In addition to RDM calculations with parameters for the lower energies from \cite{gw13}
compared with data from Refs.\,\cite{alv11,bb03,abe13}, a prediction for 5.02 TeV Pb-Pb (dashed upper curve) from Ref.\,\cite{gw16} is compared with recent
 ALICE data at 0-5\,\% centrality \cite{alice17}. A fit to the data within the five-parameter RDM is also displayed (solid curve).

Only the fragmentation sources contribute at  the lowest RHIC energy of 19.6 GeV that is shown here -- which is comparable to the highest SPS energy of 17.3 GeV --  (see dot-dashed curves), but at higher energies the gluonic source rapidly rises and
becomes the largest source of particle production at an energy of $\sim 2$ TeV, which is between energies reached at RHIC and LHC. 

I have investigated the dependence of the particle content of the three sources on center-of-mass energy per particle pair $\sqrt{s_{NN}}$ in Ref.\,\cite{gw15}. 
The gluonic source is absent for $\sqrt{s_{NN}}\lesssim 20~$GeV, see the 19.6 GeV Au-Au PHOBOS result in Figure~\ref{fig7}. At this relatively low energy, charged-hadron production arises only from the fragmentation sources which overlap in rapidity space and hence, appear like a single gaussian (``thermal'') source. The total charged-hadron production has been found experimentally to depend linearly on $\ln(s_{NN}/s_0)$, see for example central Pb-Pb NA50 data at 8.7 GeV and 17.3 GeV \cite{prino05} and low-energy Au-Au PHOBOS  results \cite{bb03}. 
\begin{figure}[t!]
\centering
\includegraphics[width=8.6cm]{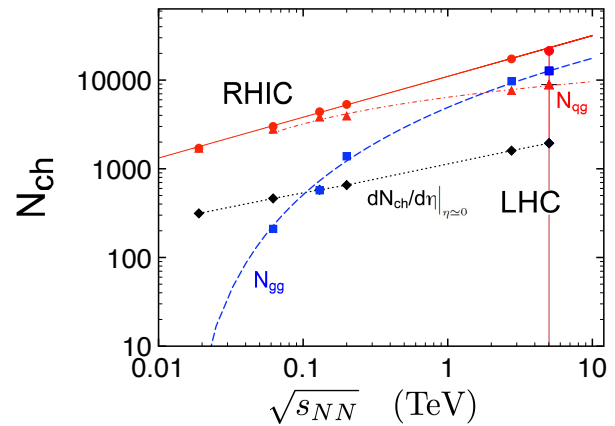}
\caption{\label{fig8} The total charged-hadron production in central Au-Au and Pb-Pb collision in the energy region 19.6 GeV to 5.02 TeV is following a power law (solid upper line), 
%$N_{tot }\propto (s_{NN}/s_0)^{0.23}$ (solid line), 
whereas the particle content in the fragmentation sources is $N_\mathrm{qg} \propto \ln{(s_{NN}/s_0})$, dash-dotted curve. The particle content in the mid-rapidity source obeys $N_{gg} \propto \ln^3{(s_{NN}/s_0)}$,
dashed curve. The energy dependence of the measured mid-rapidity yields is shown as a dotted line, with PHOBOS data \cite{alv11} at RHIC energies,
and ALICE data \cite{aamo11,ad16} at 2.76  and 5.02 TeV. The vertical line indicates 5.02 TeV.}
%\end{center}
\end{figure}

In my RDM-analysis with three sources published in Ref.\,\cite{gw15}, it has turned out that the dependence of the fragmentation sources $N_\mathrm{ch}^{qg} \propto \ln(s_{NN}/s_0)$ indeed continues at higher energies, see Figure~\ref{fig8}. In addition to what was shown in Figure\,4 of Ref.\,\cite{gw15} with an extrapolation to 5.02 TeV, now my results based on an analysis of new ALICE 5.02 TeV data \cite{ad16} are included in this figure. The gluonic source  is confirmed to have a strong energy dependence
%In Figure~\ref{fig6} the RDM analysis of 5.02 TeV Pb-Pb is added (open symbols), which is again found to fit $N_\mathrm{ch}^{qg} \propto \ln(s_{NN}/s_0)$ 
%for the fragmentation sources and 
$N_\mathrm{ch}^{gg} \propto \ln^3(s_{NN}/s_0)$. As discussed in Ref.\,\cite{gw15}, the rise of the cross section in the central distribution is driven by the growth of the gluon density at small $x$ and theoretical arguments \cite{cheu11} suggest a ln$^2 s$ asymptotic behaviour that satisfies the Froissart bound \cite{fro61}. Because the beam rapidity is $\propto \ln(s_{NN})$, the integrated yield from the gluonic source then becomes proportional to ln$^3 s$, in agreement with the phenomenological analysis, and the new 5.02 TeV data. It was mentioned in Ref.\,\cite{gw15} that there exist also
further experimental confirmations of this result at RHIC energies based on STAR data for dijet production, see \cite{tom15} and references therein.

The 5.02 TeV Pb-Pb data confirm that the sum of produced charged hadrons integrated over $\eta$ is close to a power law $N_\mathrm{ch}^{\mathrm{tot}}\propto (s_{NN}/s_0)^{0.23}$ with $s_0 = 1$ TeV$^2$ as shown in Figure~\ref{fig8} for central Au-Au and Pb-Pb collisions, upper line. At RHIC energies Busza noticed that the integrated charged-particle multiplicities scale as $\ln^2(s_{NN}/s_0)$
\cite{wbu08}, but
the energy dependence up to the maximum LHC energy of 5.02 TeV has turned to be even stronger due to the high gluon density. In Ref.\,\cite{gw15} it was shown that the midrapidity yields for central Au-Au and Pb-Pb collisions are 
\begin{equation}
\frac{dN_\text{ch}^\text{tot}}{d\eta}\biggr{|}_{\eta\simeq0}=1.15 \cdot 10^3 (s_{NN}/s_0)^{0.165}
\end{equation}
with $s_0=1~$TeV$^2$ (dotted line, data points from Phobos \cite{alv11} and ALICE \cite{aamo11,ad16}). 

 More detailed aspects of the interplay between fragmentation sources and gluonic source appear when investigating the centrality dependence of
 charged-hadron pseudorapidity distributions, as has been done in Refs.\,\cite{wobi06,sgw18} for the asymmetric systems 200 GeV d-Au and 5.02 TeV p-Pb,
 and in Refs.\,\cite{rgw12,gw13} for 2.76 TeV Pb-Pb.
  \subsection{Limiting Fragmentation at RHIC and LHC energies}
Using the RDM with both, linear and sinh-drift, we have investigated whether the limiting fragmentation conjecture is fulfilled at energies reached at RHIC and LHC \cite{kgw19}.
The significance of the fragmentation region in relativistic heavy-ion collisions had been realized when data on Au-Au collisions in the energy range $\sqrt{s_{NN}}$ ={19.6} GeV to {200}  {GeV} became available at RHIC \cite{bea02,bb03,ada06}. The pseudorapidity distributions 
%$dN_\text{ch}/d\eta$
%$(\eta=-\ln\,\tan(\theta/2))$ 
of produced charged particles for a given centrality bin scale with energy according to the limiting fragmentation (LF), or extended longitudinal scaling, hypothesis: %$dN_\text{ch}/d\eta\,'$ with $\eta\,'=\eta-y_\text{beam}$ 
Over a large range of pseudorapidities $\eta\,'=\eta-y_\text{beam}$ in the fragmentation region with the beam rapidity $y_\text{beam}$, the charged-particle pseudorapidity distribution is found to be energy independent.
%=\ln(\sqrt{s_\text{NN}}/m_p)$ with $m_p$ the mass of the proton is the beam rapidity.
%, such that $\eta\,'$ is the pseudorapidity in the rest frame of the projectile. 

The phenomenon was first shown to be present in 
p\textoverline{p} data, in a range from \num{53} up to \SI{900}{\GeV} \cite{al86}, following a
prediction for hadron-hadron and electron-proton collisions in Ref.\,\cite{benecke69}. 
%at the beam rapidity. 
With increasing collision energy the fragmentation region grows in pseudorapidity space. It can cover more than half of the pseudorapidity range over which particle production occurs.  Especially in relativistic heavy-ion collisions, the approach to a universal limiting curve is a remarkable feature of the particle production process.

\begin{figure}[t!]
\centering
\includegraphics[width=8.4cm]{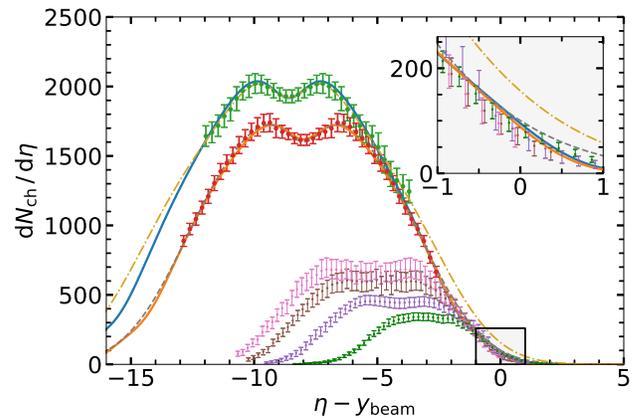}
\caption{ Comparison of the three-source RDM-distributions with linear and sinh-drift, PHOBOS data \cite{alv11},  and ALICE data \cite{abb13,alice17} with emphasis on limiting fragmentation.
From bottom to top: central  Au-Au at $\sqrt{s_\mathrm{NN}}=$19.6, 62.4, 130 and 200 GeV (RHIC), Pb-Pb at $\sqrt{s_\mathrm{NN}} = 2.76$ and 5.02 TeV. The difference between the model with sinh-drift (solid curves) and the one with linear drift (dot-dashed and dashed curves) is small, but visible in the fragmentation region. The zoom into this region shows that the RDM with sinh-drift is consistent with limiting fragmentation at RHIC and LHC energies. From Kellers and Wolschin \cite{kgw19}, where details and parameters are given.} 
\label{fig9}
\end{figure}

As discussed in Ref.\,\cite{kgw19} and references cited therein, it is an interesting question whether limiting fragmentation will persist at the much higher incident energies that are available at the LHC, namely,
$\sqrt{s_\text{NN}}$ = 2.76 and 5.02 TeV in Pb-Pb collisions. At these energies, experimental results in the fragmentation region are not available due to the lack of a dedicated forward spectrometer. If one wants to account for the collision dynamics more completely, however, this region is most interesting. We have therefore investigated in Ref.\,\cite{kgw19} to what extent limiting fragmentation can be expected to occur in heavy-ion collisions at LHC energies. The result from that investigation is summarized in Figure \ref{fig9}: Limiting-fragmentation scaling can be expected to hold at both, RHIC and LHC energies. This conclusion agrees with microscopic numerical models such as AMPT \cite{ko05}, but it disagrees with expectations from simple parametrizations of the rapidity distributions such as the difference of two Gaussians. It also disagrees  with predictions from the thermal model, which  does not explicitly treat the fragmentation sources but refers only to particles produced from the hot fireball. 
In contrast, in our approach the fragmentation sources play an essential role.
%In our phenomenological approach that confirms the validity of the LF hypothesis at LHC energies, however, the fragmentation sources play an essential role. 
Only future upgrades of the detectors would make it possible to actually test the limiting-fragmentation conjecture experimentally at LHC energies. 

 %Figure\,\ref{fig7}, Figure\,\ref{fig8}, Figure\,\ref{fig9}.
\section{Quarkonia and the QGP} 
Among the hard probes in relativistic heavy-ion collisions, the modification of quarkonia yields in the presence of the quark-gluon plasma has an outstanding role. Charmonia ($J/\psi$) suppression due to the screening of the real part of the Cornell-type quark-antiquark potential in the hot medium had initially been suggested by Matsui and Satz in 1986 as a QGP signature \cite{ms86}. It was realized later that the potential in the medium is an optical one, with the imaginary part \cite{laine07} causing dissociation of the quarkonia states in the hot medium and thus, quarkonia suppression when compared to the production rates from pp collisions at the same center-of-mass energy scaled with the number of binary collisions.
In addition to the associated collisional damping widths of the quarkonia states, thermal gluons can dissociate these states, and the gluon-induced dissociation widths \cite{bgw12}  can be treated separately from the damping \cite{ngw14}. An important role is played by the reduction of feed-down in the heavy-ion case as compared to pp, because feed-down from the higher to the lower states is hindered if the higher states are screened away, or depopulated. If the medium contains a large number of heavy quarks as is the case for charm quarks in Pb-Pb collisions at LHC energies, statistical recombination cannot be neglected at sufficiently low transverse momentum, and there is an interplay of dissociation and recombination. There is meanwhile a large number of publications and reviews about charmonia physics in relativistic heavy-ion collisions available \cite{qm19}.
\begin{figure}
	\centering
	\includegraphics[width=7cm]{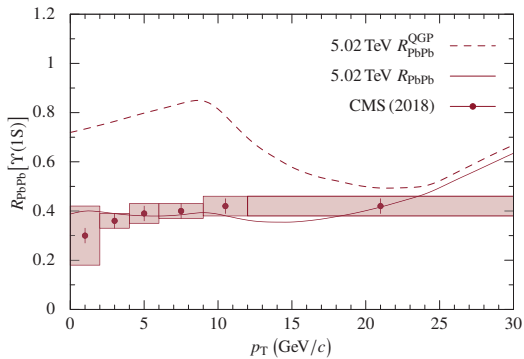}
	\caption{Transverse-momentum dependence of the suppression factor $R_\text{AA}(\Upsilon(1S))$ for the spin-triplet ground state 
		  in minimum-bias Pb-Pb~collisions at $\sqrt{s_{NN}} = 5.02$ TeV. The (upper) dashed curve shows the suppression in the hot medium, the (lower) solid curve the suppression including reduced feed-down, which is important for the ground state, but not for excited states.  The theoretical prediction is from 
		  Ref.\,\cite{hnw17}, data are from CMS \cite{cms19}. }
\label{fig10}
\end{figure}
\begin{figure}
	\centering
	  %\begin{minipage}[t]{0.41\textwidth}
    \includegraphics[width=8cm]{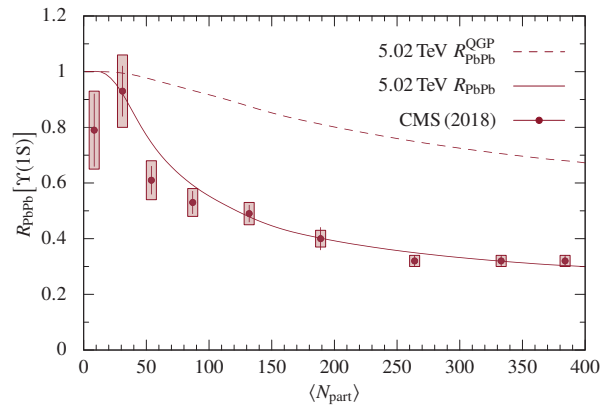}
    \caption{Centrality-dependent suppression factor $R_\text{AA}(\Upsilon(1S))$ in Pb-Pb~collisions at $\sqrt{s_{NN}} = 5.02$ TeV (solid line, from Ref.\,\cite{hnw17})  together with data from CMS (dots, $|y| < 2.4$, Ref.\,\cite{cms19}),  as function of the number of participants $\Npart$ (averaged over centrality bins).
		The suppression in the QGP-phase is 
		%$R^{QGP}_{AA}(\Upsilon(1S))$ (
		the dashed curve, the solid curve includes reduced feed-down. Theoretical prediction from Ref.\,\cite{hnw17}.}
		\label{fig11a}
		\end{figure}
 % \end{minipage}
  %\hfill
 % \begin{minipage}[t]{0.42\textwidth}
 \begin{figure}
	\centering
    \includegraphics[width=8cm]{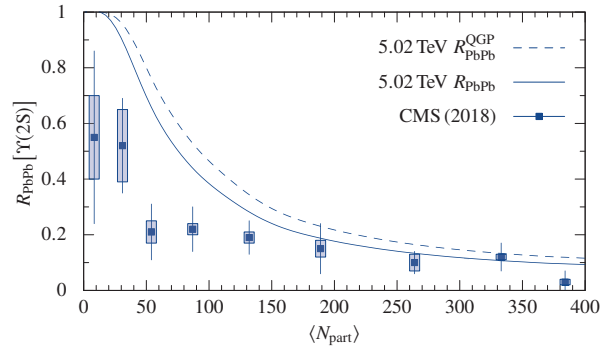}
    \caption{Suppression factor for the first excited spin-triplet state $R_\text{AA}(\Upsilon(2S))$ in Pb-Pb~collisions at $\sqrt{s_{NN}} = 5.02$ TeV (solid line) together with data from CMS. The suppression factor $R_\text{AA}(\Upsilon(2S))$ in the QGP-phase (dashed) accounts for most of the calculated total suppression (solid) for the $\Upsilon(2S)$.
		Theoretical predictions are from Hoelck, Nendzig, Wolschin \cite{hnw17}, CMS data are from Ref.\,\cite{cms19}.}
		\label{fig11b}
 % \end{minipage}
	%\includegraphics[scale=0.8]{fig11a}
%	\caption{
%		Left: Centrality-dependent suppression factor $R_\text{AA}(\Upsilon(1S))$ in Pb-Pb~collisions at $\sqrt{s_{NN}} = 5.02$ TeV (solid line)  together with data from CMS (dots, $|y| < %2.4$),  as function of the number of participants $\Npart$ (averaged over centrality bins).
%		The suppression factor $R^{QGP}_{AA}(\Upsilon(1S))$ in the QGP-phase is shown as dashed (upper) curve, the solid (lower) curve includes reduced feed-down.\\
		%A prediction of $\RAAYnS{1}$ for $\sNN = \WithUnit{5.02}{TeV}$ (dash-dotted line, and dotted line for corresponding $\RAAQGP$) as a function of centrality is plotted for %a %formation time $\tauFnl = \WithUnit{0.4}{fm/$c$}$.
%		Right: Suppression factor for the first excited spin-triplet state $R_\text{AA}(\Upsilon(2S))$ in Pb-Pb~collisions at $\sqrt{s_{NN}} = 5.02$ TeV (solid line) together with data from %CMS. The suppression factor $R_\text{AA}(\Upsilon(2S))$ in the QGP-phase (dashed) accounts for most of the calculated total suppression (solid) for the $\Upsilon(2S)$.
%		Theoretical predictions are from Hoelck, Nendzig, Wolschin \cite{hnw17}, CMS data are from Ref.\,\cite{cms19}.}
		%Most of the suppression for excited states occurs in the QGP-phase. Additional suppression mechanisms are required in particular for peripheral collisions.}
	%\label{fig11}
\end{figure}

Bottom quarks are about three times heavier than charm quarks, have a correspondingly smaller production cross section, and hence, are less abundant even at LHC energies.
Statistical recombination is therefore less important, and expansions in terms of $(1/m)$ are more precise. Consequently, bottomonia provide a cleaner probe of the QGP properties such as the initial central temperature, and have  been investigated in detail both theoretically and experimentally. The bottom quarks are produced on a very short time scale of 0.02 fm/$c$ in the initial stages of the collision, before the QGP of light quarks and gluons is actually being produced. The formation time of bottomonia states is larger, in the range $\tau_\text{F}\simeq 0.3-0.6$ fm/$c$. It is less precisely known, may differ for the individual states such as 
$\Upsilon(1S,2S,3S)$ and $\chi_b(1P,2P,3P)$, and could depend on the temperature of the emerging QGP, which would further enlarge it \cite{ko15}. Since the spin-triplet $\Upsilon(1S)$ state is particularly stable with a binding energy of $\simeq 1.1$ GeV, it has a sizeable probability to survive as a color-neutral state in the colored hot quark-gluon medium of light quarks and gluons that is created in a central Pb-Pb collision at LHC energies, even at initial medium temperatures of the order of 400 MeV or above.

There exists a considerable literature on the dissociation of quarkonia, in particular of the $\Upsilon$~meson \cite{CMS-2012,ab14,ada14}, in the hot quark-gluon medium; see Ref.\,\cite{an16} and references therein for a review. In minimum-bias Pb-Pb-collisions at LHC energies of $\sqrt{s_{NN}} = 5.02$ {TeV} in the midrapidity range, the strongly bound 
$\Upsilon(1S)$-state is found to be suppressed down to about 38\,\% as compared to the expectation from scaled pp collisions at the same energy. The $\Upsilon(1P)$ state has a smaller binding energy and is even more suppressed, down to 12\,\% \cite{cms19}. 

Various theoretical approaches such as Refs.\,\cite{em12,striba12,peng11,song12}, and their recent updates to higher energy, are available that allow for an interpretation of the data.  In the next section, results of our model are reported that aims to account for the previous Pb-Pb results at 2.76 TeV
% in terms of screening, damping, gluodissociation and reduced feed-down,
and to predict results for the higher energy of 5.02 TeV, which are then compared to the data that have meanwhile become available.
\subsection{$\Upsilon(1S,2S)$ suppression in Pb-Pb at LHC energies}
In Refs.\,\cite{ngw13,ngw14,hnw17} we have devised a model that accounts for the screening of the real part of the potential, the gluon-induced dissociation of the various bottomonium states in the hot medium (gluodissociation), and the damping of the quark-antiquark binding due to the presence of the medium which generates an imaginary part of the temperature-dependent potential.
Screening is less important for the strongly bound $\Upsilon(1S)$ ground state, but it is relevant for the $\bb$ excited states, and also for all $\cc$ bound states.

Due to screening and depopulation of the excited states in the hot medium, the subsequent feed-down cascade towards the $\Upsilon(1S)$ ground state differs considerably from what is known based on pp collisions. The LHCb collaboration has measured a feed-down fraction of $\Upsilon(1S)$ originating  from $\chi_b(1P)$ decays in pp collisions at $\sqrt{s} = {7}$ {TeV} of 20.7\,\% \cite{lhcb12}, and the total feed-down from excited states to the ground state is estimated to be around 40\,\% \cite{gman18} at LHC energies. If feed-down was 
completely absent because of screening and depopulation of excited states in the hot medium, a modification factor $R_\text{AA}(\Upsilon(1S))\simeq 0.6$ would thus result, whereas the measured modification factor of the $\Upsilon(1S)$ state  in minimum-bias Pb-Pb collisions at 2.76 TeV is $R_\text{AA}(\Upsilon(1S))=0.453$ $ \pm 0.014$ (stat)$\pm 0.046$ (syst) 
\cite{cms17}, and at 5.02 TeV  $R_\text{AA}(\Upsilon(1S))=0.378\pm 0.013$ (stat)$\pm 0.035$
%$R_\text{AA}(\Upsilon(1S))=0.378\pm  0.013$  (stat)$\pm  0.035 $ 
(syst)  \cite{cms19}. Hence, there clearly exist in-medium suppression mechanisms for 
the strongly bound $\Upsilon(1S)$ state which we aim to account for in detail, together with the suppression of the excited states, and the reduced feed-down.

In our model calculation \cite{hnw17}, we thus determine the respective contributions from in-medium suppression, and from reduced feed-down for the 
$\Upsilon(1S)$  ground state, and the $\Upsilon(2S)$ first excited state in Pb-Pb collisions at both LHC energies, 2.76 TeV and 5.02 TeV.
The $p_\text{T}$-dependence and the role of the relativistic Doppler effect on the measured transverse-momentum spectra are considered.
For the $\Upsilon(2S)$  state, the QGP effects are expected to be much more important than reduced feed-down. 
We compare in Ref.\,\cite{hnw17} with centrality-dependent CMS data~\cite{CMS-2012,cms17} for the $\Upsilon(1S)$ and $\Upsilon(2S)$  states in ${2.76}$ {TeV} Pb-Pb collisions, and predict the $p_\text{T}$- and centrality-dependent suppression at the higher LHC energy of $\sqrt{s_{NN}}= {5.02}$ {TeV}. The predictions at 5.02 TeV are compared with recent CMS data in Fig.\,\ref{fig10} for the transverse-momentum dependence, Fig.\,\ref{fig11a} for the centrality dependence of $\Upsilon(1S)$, and Fig.\,\ref{fig11b} for the centrality dependence of $\Upsilon(2S)$.
%and is compared in this note with CMS data \cite{cms19}.

For symmetric systems such as Au-Au at RHIC or Pb-Pb at LHC, we do not include an explicit treatment of CNM effects such as shadowing in the present study. These are, however, important in asymmetric collisions such as p-Pb where most of the system remains cold during the interaction time, and we have considered them in our corresponding calculations \cite{dhw19}.
% shown at the end of this note.
%In symmetric systems at RHIC and LHC energies, however, the CNM effects such as shadowing are likely less important and moreover, expected to be very similar for ground and %excited states.
Statistical recombination of the heavy quarks following bottomonia dissociation is disregarded: Although this is certainly a relevant process in the 
$J/\psi$ case, the cross section for $\Upsilon$ production is significantly smaller.

The anisotropic expansion of the hot fireball is accounted for using hydrodynamics for a perfect fluid that includes transverse expansion. Such a simplified nonviscous treatment \cite{ngw14,hnw17} 
of the bulk evolution appears to be tolerable because conclusions on the relative importance of the in-medium suppression versus reduced feed-down are not expected to depend much on the details of the background model. When calculating the in-medium dissociation, we consider the relativistic Doppler effect that arises due to the relative velocity of the bottomonia with respect to the expanding medium. It leads to more suppression at high $p_\text{T}$, and to an overall rather flat dependence of $R_\text{AA}$ on $p_\text{T}$.

Our predictions for the $p_\text{T}$-dependent $\Upsilon$-suppression in 5.02 TeV Pb-Pb collisions are shown together with recent CMS data \cite{cms19} in 
Fig.\,\ref{fig10}; see the caption for details. The in-medium modification factor (dashed) first rises at small transverse momentum, because escape from the hot zone becomes more likely with increasing $p_\text{T}$, but then falls off when the increase in effective temperature becomes more pronounced. For the $\Upsilon(1S)$ state, a substantial fraction of the suppression, in particular at low $p_\text{T}$, is due to reduced feed-down, solid curve. The corresponding centrality-dependent suppression (integrated over $p_\text{T}$) is shown in Fig.\,\ref{fig11a} to be in agreement with the data \cite{cms19} for the $\Upsilon(1S)$ state.
Related ALICE data at more forward rapidities $2.5<y<4$ are roughly consistent within the error bars \cite{alice19b}.
The suppression of the $\Upsilon(2S)$ state in Fig.\,\ref{fig11b}  is mostly in-medium, with only a small contribution due to reduced feed-down. The prediction shows less suppression than the data in peripheral collisions. We have shown in Ref.\;\cite{hgw17} that the extra suppression of the loosely bound $\Upsilon(2S)$ state is most likely \it{not} \rm due to the strong electromagnetic fields in more peripheral collisions. Hence, the origin of this effect is presently unknown.

\begin{figure}
	\centering
	\includegraphics[width=8cm]{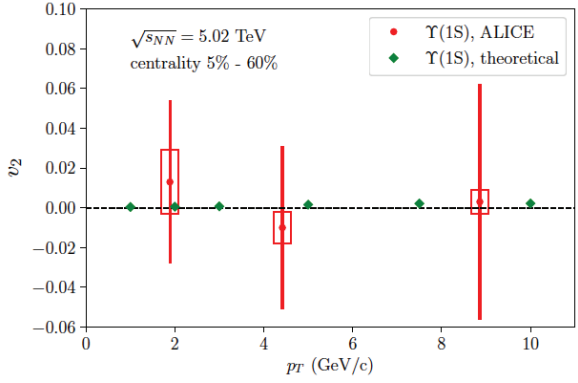}
	\caption{The ellipticity $v_2(p_\text{T})$ of the $\Upsilon(1S)$ momentum distribution in  $5-60\%$ Pb-Pb
		  collisions at $\sqrt{s_\text{NN}}=5.02$TeV as calculated from anisotropic escape is consistent with zero, green symbols. (From Fritsch, BSc thesis HD 2020, unpublished. 
		  Reproduced with permission.) The data (red) are from ALICE 
		  \cite{alice19a}. 
		  }
\label{fig12}
\end{figure}

It is of considerable interest to determine if the bottomonia distributions in more peripheral collisions become anisotropic, as has been found for produced particles in general. The quadrupole part of the momentum anisotropy is due to the almond-shaped spatial anisotropy of the overlap region, which translates to momentum space. It is more pronounced for lighter mesons
such as pions, and can be quantified by the ellipticity $v_2$ of the momentum distribution in a Fourier decomposition of the experimentally determined, event-averaged particle distribution \cite{hesne13} 
\begin{equation}
\frac{d\langle N \rangle}{d\phi}=\frac{\langle N \rangle}{2\pi}\left(1+2\sum_{n=1}^{\infty}\langle v_n \rangle \cos[n(\phi-\langle \psi_n \rangle)]\right)
\end{equation}
with the azimuthal angle $\phi$, the mean flow angle $\langle \psi_n \rangle$,  and $\langle N \rangle$ 
the mean number of particles of interest per event (charged hadrons or
identified particles of a specific species). The flow planes are, however, not experimentally known, and hence, the anisotropic flow coefficients are
obtained using azimuthal angular correlations between the observed particles. The experimentally reported anisotropic flow coefficients from two-particle correlations can then be obtained as the
root-mean-square values, $v_n\{2\} \equiv v_n\equiv\sqrt{\langle v_n^2 \rangle}$, and the flow coefficients are being measured not in individual events, but in centrality classes.

Whereas for charged hadrons the flow coefficients have been measured very precisely \cite{hesne13} with $v_2$-values up to $20-30\%$ for pions, kaons, and antiprotons and maxima near $p_\text{T}\simeq 3$ GeV/$c$, this is more difficult for quarkonia due to the much smaller production rates. For the charmonium ground state, the ellipticity coefficient at  forward rapidity ($2.5<y<4$) in the centrality class $5-60\%$ is $v_2\simeq3-8\%$, depending on $p_\text{T}$ \cite{alice19a}, implying that $J/\psi$ shows elliptic flow, albeit on a smaller scale due to the larger mass. 
\begin{figure}
	\centering
	\includegraphics[width=4.6cm]{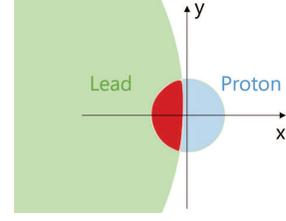}
	\caption{Overlap (red) of the thickness functions in the transverse plane for lead (green) and proton (blue). (From Dinh, MSc thesis HD 2019, unpublished. Reproduced with permission.)  }
\label{fig13}
\end{figure}
\begin{figure}
	\centering
	\includegraphics[width=8.0cm]{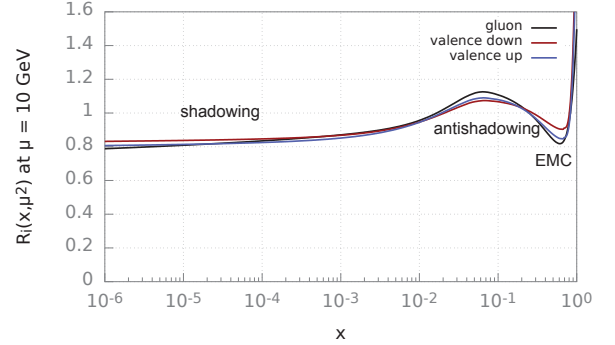}
	\caption{Modification of the nuclear PDFs EPPS16 \cite{es17} for gluons, up- and down-quarks as function of the momentum fraction $x$ at $10^{-6} < x  \le 1$: Shadowing at $x \le 0.02$, antishadowing at $0.02<x<0.3$. (From Dinh, MSc thesis HD 2019, unpublished. Reproduced with permission.)  }
\label{fig14}
\end{figure}
 \begin{figure}[b]
	\centering
	\includegraphics[width=8.0cm]{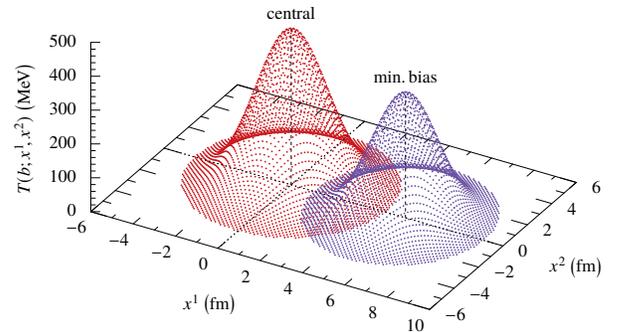}
	\caption{Initial temperature profiles of the hot QGP generated in p-Pb collisions at $\sqrt{s_{NN}} = 8.16$ TeV as functions of the transverse coordinates $(x^1,x^2)$ at two centralities: Central collisions with $N_\text{coll} \simeq 15.6$, left, and minimum-bias collisions with $N_\text{coll} \simeq 7$, right.  From Ref.\,\cite{dhw19}.}
\label{fig15}
\end{figure}
\begin{figure*}
	\centering
	\includegraphics[width=10cm]{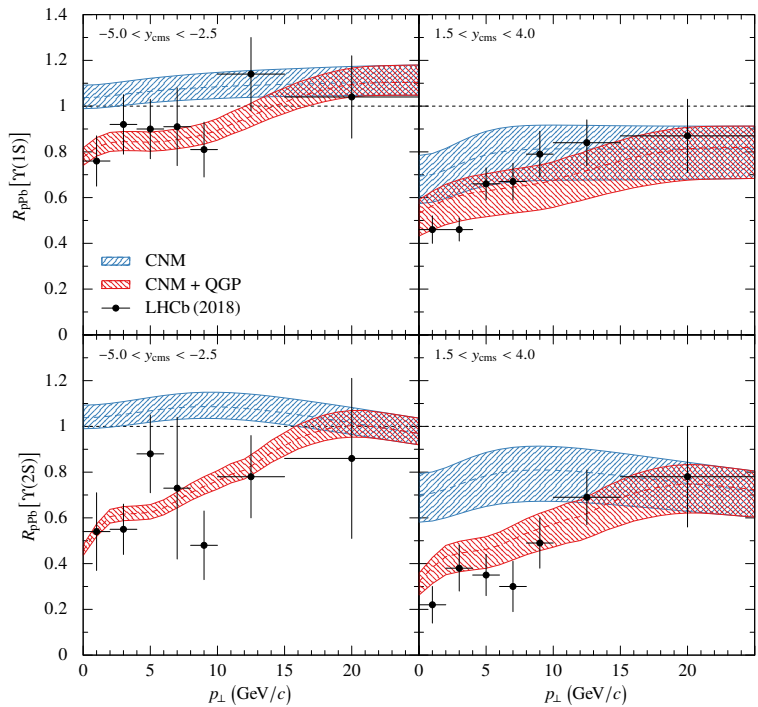}
	\caption{Calculated transverse-momentum-dependent nuclear modification factors $R_\text{p-Pb}$ for the $\Upsilon(1S,2S)$~spin-triplet ground and first excited state in p-Pb~collisions at $\sqrt{s_{NN}} = 8.16$ TeV with LHCb data 
	\cite{lhcb18} in the backward (Pb-going, left) and forward (p-going, right) region, for minimum-bias centrality.
		Results for CNM effects that include shadowing, energy loss, and reduced feed-down (dashed curves, blue) are shown together with calculations that incorporate also QGP effects (solid curves, red).
		%In both cases, the feed-down cascade is included.
		The error bands result from the uncertainties of the parton distribution functions that enter the model. Calculations from Ref.\,\cite{dhw19}.
 }
\label{fig16}
\end{figure*}
\begin{figure}
	\centering
	\includegraphics[width=7.6cm]{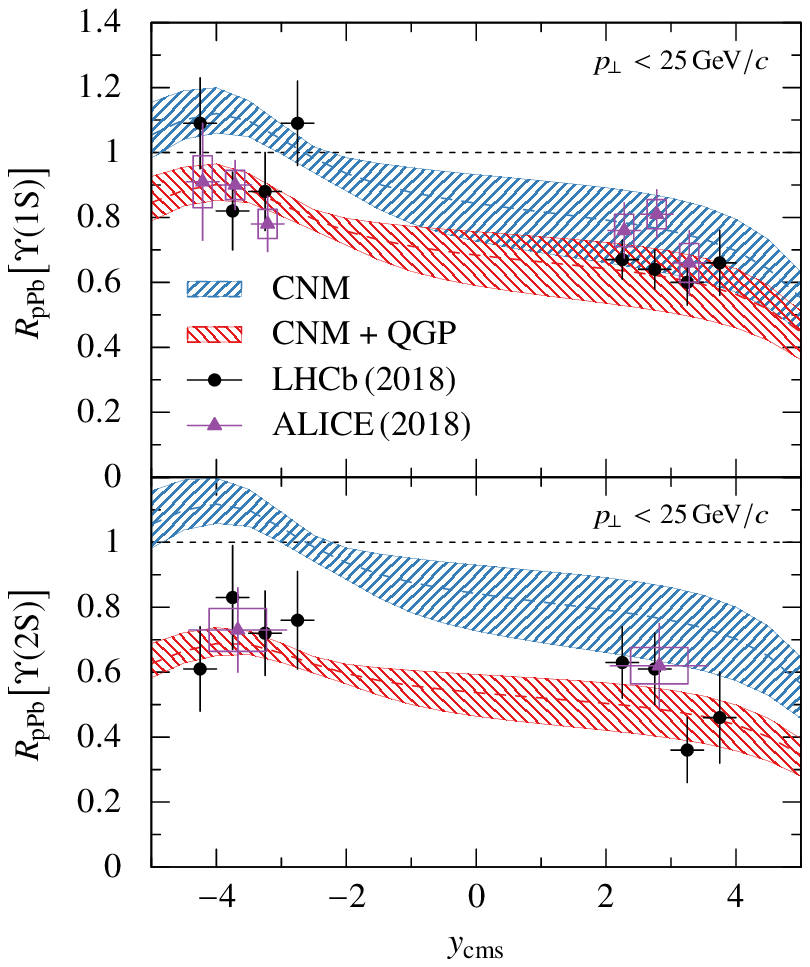}
	\caption{Calculated rapidity-dependent nuclear modification factors $R_\text{p-Pb}$ for the $\Upsilon(1S)$ (top) and $\Upsilon(2S)$~state (bottom) in p-Pb~collisions at  
	$\sqrt{s_\text{NN}}=8.16$ TeV with  preliminary ALICE data \cite{alicep18} (triangles) and LHCb data
	 \cite{lhcb18} (circles).
		Results for CNM effects that include shadowing, energy loss, and reduced feed-down (dashed curves, blue) are shown together with calculations that incorporate also QGP effects (solid curves, red).
		%In both cases, the feed-down cascade is included.
		The error bands result from the uncertainties of the parton distribution functions that enter the calculations. From Dinh, Hoelck, Wolschin \cite{dhw19}.
 }
\label{fig17}
\end{figure}
It may, therefore, appear possible that even bottomonium exhibits flow, following the mass ordering of lighter particles resulting from a collective
expansion of the medium \cite{rey19}. Indeed, the large statistical error bars on the presently available bottomonium data for $v_2$ \cite{alice19a} do not yet exclude such a possibility,
although it is quite doubtful whether this meson -- which is about three times heavier than charmonium -- flows with the expanding hot medium. Instead, the $\Upsilon(1S)$ is expected to essentially maintain its trajectory in the hot QGP, unless it is dissociated. Still, its momentum distribution in more peripheral collisions may exhibit a finite $v_2$ due to the anisotropic escape from the fireball, because the path length from $\Upsilon$ formation to escape in the transverse plane depends on the azimuthal angle. This mechanism had already been suggested for $J/\psi$ by Wang and Yuan at SPS and RHIC energies \cite{wang02}, and has been used in Ref.\,\cite{bbjs19} for the bottomonium states in non-central 2.76 TeV Pb-Pb. There, the maximum of $v_2(\Upsilon(1S))$ including feed-down contributions is found to be below $1\%$ in the $40-50\%$ centrality class.
We have performed a corresponding calculation within our model in the $5-60\%$ centrality class and compare the $p_\text{T}$ dependence with the available ALICE data \cite{alice19a} , see Figure\,\ref{fig12}. The anisotropy is very small, compatible with zero. For more definite conclusions, one has to wait for a reduction of the experimental uncertainties in run 3.

%Ref.\,\cite{fela18}
% In the next section, comparisons of the model pre
%Figure\,\ref{fig10}, Figure\,\ref{fig11a}, Figure\,\ref{fig11b},  Figure\,\ref{fig12}
\subsection{$\Upsilon(1S,2S)$ modification in p-Pb at $\sqrt{s_{NN}} = 8.16$ TeV }

Regarding bottomonia in asymmetric collisions, p-Pb at $\sqrt{s_{NN}} = 8.16$ TeV has been investigated experimentally by the LHCb 
\cite{lhcb18} and ALICE \cite{alice20} collaborations, 
and  cold nuclear matter predictions had been published by a group of theorists \cite{alba18}. Clearly, CNM effects are much more relevant than in symmetric systems, because the bulk of the hadronic matter remains cold during the interaction, see Fig.\,\ref{fig13}. The most relevant CNM effect is the modification of the parton distribution functions in the nuclear medium, which have been studied by many authors. A typical result for the PDF modifications with shadowing at small values of Bjorken-$x$, and antishadowing at intermediate $x$-values as obtained with EPPS16 \cite{es17} is shown in Fig.\,\ref{fig14}. Shadowing causes a reduction of the $\Upsilon(nS)$ yields in p-Pb as compared to scaled pp, whereas antishadowing results in an enhancement. Shadowing is somewhat more pronounced if one, in addition, considers coherent energy-loss mechanisms in the cold medium. Still, these are not sufficient to interpret the available data in terms of CNM effects, as becomes obvious from direct comparisons, in particular, for the $\Upsilon(2S)$ state.

There is, however, a spatially small hot zone (fireball) with an initial central temperature that is comparable to the one in a symmetric system (Fig.\,\ref{fig15}), and during its expansion and cooling, it contributes to bottomonia dissociation in regions where the temperature remains above the critical value. We have investigated the respective cold-matter and hot-medium effects on 
$\Upsilon$-dissociation in 8.16 TeV p-Pb collisions in Ref.\,\cite{dhw19}. 
Representative results from this work are shown in
Fig.\,\ref{fig16} for the transverse-momentum dependence, and Fig.\,\ref{fig17} for the rapidity dependence. The plots show CNM (blue, upper bands)  and CNM plus QGP (red, lower bands) effects on the $\Upsilon(1S)$ and $\Upsilon(2S)$ yields in 8.16 TeV p-Pb collisions at the LHC. The transverse-momentum dependence in the backward direction (top) shows enhancement
due to antishadowing when only the CNM effects are considered, whereas the data for $\Upsilon(1S)$ are clearly suppressed at $p_\text{T} <10$ GeV/$c$ and for $\Upsilon(2S)$ at all measured $p_\text{T}$ values. This discrepancy is cured through the consideration of the momentum-dependent dissociation in the QGP, as shown in our cold-matter plus hot-medium calculation (red).

The forward/backward asymmetric shape of the nuclear modification factors as functions of rapidity (Fig.\,\ref{fig17}) arises from the different cold-matter effects in the forward and backward regions
(in particular, shadowing/antishadowing of the parton distribution functions, but also energy loss in the relatively cold medium). The additional suppression due to the dissociation in the hot fireball is again shown in the lower (red) curves, which are in better agreement with the data for the $\Upsilon(1S)$ ground state not only in the backward, but also in the forward direction. The substantial role of the hot-medium effects is even more pronounced for the $\Upsilon(2S)$  first excited state, where the CNM calculation shows enhancement in the backward region, whereas the full calculation with in-medium dissociation displays a suppression down to about 70\%  -- in agreement with the LHCb data 
\cite{lhcb18} and the ALICE data point \cite{alice20}.

There have been attempts to explain the discrepancy between CNM calculations and data for the $\Upsilon(nS)$ suppression in p-Pb in terms of interactions with comoving hadrons, in particular, pions \cite{fela18}. We have not included this process in our calculations -- initially on the grounds that interactions of the bottomonia states with comovers were found to be unimportant at LHC energies in the work of Ko et al.\,\cite{ko01} about $\Upsilon$ absorption in hadronic matter. Probably one eventually has to consider both, comover interactions plus suppression in the hot QGP zone in order to fully understand the $\Upsilon$ modification data in asymmetric systems.
%Fig.\,\ref{fig13}, Fig.\,\ref{fig14}, Figure\,\ref{fig15}, Figure\,\ref{fig16}, Figure\,\ref{fig17}.
%\cite{ko01}
% and RDM parameters as in table 1. 
%The solid curve is a $\chi^2$-minimization based on the three-sources RDM with respect to the preliminary %ALICE data from \cite{to11}
%that takes the limiting fragmentation hypothesis into account, see text. Error bars have been symmetrized %using the larger branches. The corresponding 
%RDM parameters are given in Table~\ref{tab1}, and ~2. 
%, with the dashed curve
%arising from gluon-gluon collisions in the fireball, the dash-dotted curve from valence quark-gluon events in the Pb-like region, and the dotted curve correspondingly in the proton-like %region.} 
\section{Conclusions}
This article presents aspects of relativistic heavy-ion collisions with an emphasis on energies reached at the Relativistic Heavy Ion Collider RHIC and the Large Hadron Collider LHC. It does not attempt to be a review of the field, which is available in recent textbooks and in the proceedings of Quark Matter conferences such as Ref.\,\cite{qm19}. Instead, a specific phenomenological viewpoint with experimental data as a guiding principle is taken, but also QCD-based and nonequilibrium-statistical arguments are considered. The rapid local equilibration of gluons and quarks in the initial stages of a relativistic heavy-ion collision is being modeled through exact analytical solutions of a nonlinear diffusion equation. On a similar time scale, stopping is accounted for in a QCD-inspired model, which is also incorporated in a nonequilibrium-statistical approach to compute the time evolution from the initial to the measured distribution functions of net protons. The production of charged mesons such as pions, kaons, and antiprotons is discussed in a phenomenological three-source relativistic diffusion model that emphasises the importance of the fragmentation distributions in addition to the usual fireball source. These are also shown to be essential in the phenomenon of limiting fragmentation, which had been confirmed experimentally at RHIC energies, and turns out to be consistent with the present LHC energies. The investigation of the dissociation of quarkonia in the QGP provides insights into the QGP properties, including an indirect measurement of its initial central temperature before the anisotropic expansion sets in. In asymmetric systems such as p-Pb at the current maximum LHC energy, the interplay of cold-matter and hot-medium effects has been studied, achieving a detailed understanding of the available data from the Large Hadron Collider. The more precise measurements from the forthcoming run 3 at the LHC are expected to provide deeper insights.

%\vspace{.2cm}

%\funding{This research received no external funding.}
%%%%%%%%%%%%%%%%%%%%%%%%%%%%%%%%%%%%%%%%%%
\acknowledgments{I am grateful to members of UHD's Multiparticle Dynamics Group for discussions and collaborations, which resulted in common publications that are quoted in the references. Special thanks go to the Heidelberg students  Viet Hung Dinh (now Orsay), Johannes H\"olck, Benjamin Kellers, Niklas Rasch, Philipp Schulz, and Alessandro Simon.}

%%%%%%%%%%%%%%%%%%%%%%%%%%%%%%%%%%%%%%%%%%
%\conflictsofinterest{The author declares no conflict of interest.}

%\acknowledgments
%Discussions with Niklas Rasch in the course of his BSc thesis and Ref.\,\cite{rgw20} are gratefully acknowledged.
%\newpage
%\reftitle{References}
%\bibliographystyle{plain}
\bibliography{gw_20}
%\end{thebibliography}

\end{document}